\documentclass[usenatbib,usegraphics]{mn2e}                 %

\def\kms{km\,s$^{-1}$}

\def\M{M$_{\odot}$}

\def\bv{B$-$V}

\def\ebv{E(B$-$V)}
\def\rv{$R_{V}$}
\def\dm15{$\Delta m_{15}$(B)}
\def\psec{\ensuremath{\,.\!\!^{s}}}
\def\parcsec{\ensuremath{\,.\!\!\arcsec}}

\def\CII{C\,{\sc ii}}
\def\OI{O\,{\sc i}}

\def\NaI{Na\,{\sc i}}

\def\SII{S\,{\sc ii}}
\def\SIII{S\,{\sc iii}}

\def\SiII{Si\,{\sc ii}}
\def\SiIII{Si\,{\sc iii}}
\def\MgII{Mg\,{\sc ii}}
\def\CaII{Ca\,{\sc ii}}
\def\FeII{Fe\,{\sc ii}}
\def\FeIII{Fe\,{\sc iii}}
\def\CoII{Co\,{\sc ii}}

\def\NiII{Ni\,{\sc ii}}

\def\HeI{He\,{\sc i}}

\def\ergs{\,(erg\,s$^{-1}$)}

\input{psfig.tex}

\title[Supernova 2003cg]
{Anomalous extinction behaviour towards the Type Ia SN 2003cg.}

\author[Elias-Rosa et al.]
{N. Elias-Rosa$^{1,2}$, S. Benetti$^{1}$, E. Cappellaro$^{1}$, M.
Turatto$^{1}$, P. A. Mazzali$^{3,4}$,
\newauthor
F. Patat$^{5}$, W.P.S. Meikle$^{6}$, M. Stehle$^{3}$, A.
Pastorello$^{3}$, G. Pignata$^{7}$,
\newauthor
R. Kotak$^{5}$, A. Harutyunyan$^{1}$, G. Altavilla$^{8}$, H.
Navasardyan$^{1}$, Y. Qiu$^{9}$,
\newauthor
M. E. Salvo$^{10}$ and W. Hillebrandt$^{3}$.
\\
$^1$INAF - Osservatorio Astronomico di Padova, vicolo
dell'Osservatorio 5, I-35122 Padova, Italy \\
$^2$Universidad de La Laguna, Av Astrof\'isico Fransisco S\'anchez
s/n, E-38206. La Laguna, Tenerife, Spain\\
$^3$Max-Planck-Institut f\"{u}r Astrophysik, Karl-Schwarzschild-Str. 1, D-85748 Garching bei M\"{u}nchen, Germany\\
$^4$INAF - Osservatorio Astronomico di Trieste, via Tiepolo 11, I-34131 Trieste, Italy\\
$^5$European Southern Observatory, Karl-Schwarzschild-Str. 2, D-85748 Garching bei M\"{u}nchen, Germany\\
$^6$Astrophysics Group, Imperial College London, Blackett Laboratory, Prince Consort Road, London, SW7 2AZ, U.K.\\
$^7$Departamento de Astronom\'{i}a y Astrof\'{i}sica, Pontificia Universidad Cat\'{o}lica, Chile\\
$^8$Departament d'Astronomia i Meteorologia, Universitat de Barcelona, Mart\'{i} i Franqu\`{e}s 1, E-08028 Barcelona, Spain\\
$^9$National Astronomical Observatories, Chinese Academy of Sciences, 100012 Beijing, China\\
$^{10}$Australian National University, Mount Stromlo Observatory, Cotter Road, Weston ACT 2611 Canberra, Australia\\
}

\date{Received ................; accepted ................}

\begin{document}

\maketitle

\begin{abstract}
We present optical and near-infrared photometry and spectroscopy
of the Type~Ia SN 2003cg, which exploded in the nearby galaxy NGC
3169. The observations cover a period between -8.5 and +414 days
post-maximum. SN~2003cg is a normal but highly-reddened Type~Ia
event. Its B magnitude at maximum $B_{max}$=15.94$\pm$0.04 and
$\Delta m_{15}(B)_{obs}$ = 1.12$\pm$0.04 ($\Delta
m_{15}(B)_{intrinsic}$ = 1.25$\pm$0.05). Allowing \rv~to become a
free parameter within the Cardelli et al. (1989) extinction law,
simultaneous matches to a range of colour curves of normal SNe~Ia
yielded \ebv=1.33$\pm$0.11, and \rv=1.80$\pm$0.19. While the value
obtained for \rv~is small, such values have been invoked in the
past, and may imply a grain size which is small compared with the
average value for the local ISM. \end{abstract}


\section{Introduction} \label{int}

Type Ia Supernovae (hereafter SNeIa) became very popular in the
last decade because of their role in determining the geometry of
the Universe. Because of their high luminosity and relatively
small dispersion at maximum, SNeIa are the most accurate
cosmological distance indicators currently available. Systematic
studies of these events at high redshifts (z $\sim$ 0.3 - 1.6),
together with cosmic microwave background data and cluster masses
and abundances, have provided strengthening evidence that the
expansion of the Universe began to accelerate at half its present
age. This finding is commonly related to a positive cosmological
constant $\Lambda$ (\citealt{riess98}; \citealt{perlmutter99}) and
is generally attributed to a new form of dark energy
\citep{knop03}.  \\

In spite of the importance of these results, caveats exist in the
cosmological use of SNe~Ia: 1) that the properties of high--z
SNe~Ia are the same as those of their present-time counterparts
is, to some extent, an assumption. Observational confirmation is
highly desirable; 2) the use of SNe Ia as cosmological distance
indicators relies on an empirical relation between the luminosity
and the light curve shape which is only partly understood; 3) the
progenitor systems of SNe Ia have not yet been unambiguously
identified, nor have and the explosion mechanisms been clearly
determined; 4) the extent to which observations of SNe~Ia are
affected by dust extinction is still unresolved.

The theoretical and observational investigation of local SNe~Ia is
the main motivation of the European Supernova Collaboration
(ESC)\footnote{http://www.mpa-garching.mpg.de/$\sim$rtn/}. This
collaboration was born under the conviction that the only way to
make a decisive jump in understanding the physics of SNe~Ia is to
compare realistic models of thermonuclear explosions (and the
emerging radiation) with a complete and detailed set of
homogeneously-acquired observations. The first two years of the
ESC produced excellent results, with fourteen well-monitored
targets, e.g. \cite{benetti04}, \cite{pignata04} and
\cite{kotak05}.\\

In this paper, we present optical and infrared photometry and
spectroscopy of SN~2003cg, the fourth supernova monitored by the ESC.
It is exceptional in that it is heavily reddened. \\

\section{Observations} \label{obs}
\begin{figure*}~
\centering \psfig{figure=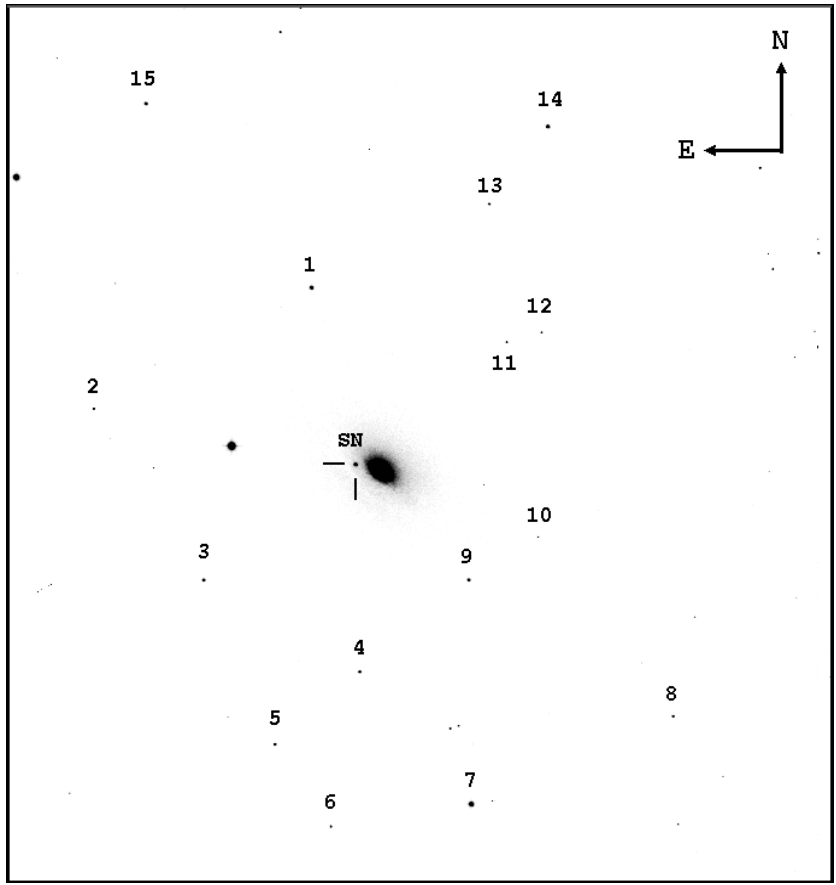,width=11cm,height=11.7cm}
\caption{V band image of SN 2003cg in NGC 3169 taken with the
ESO/MPI 2.2m+WFI on 2003 April 03 (FoV $\sim 8'.5\times9'.0$).
Local sequence stars are indicated (cf. Table \ref{tabla_seq}).}
\label{fig_seq}
\end{figure*}

SN 2003cg was discovered on 2003 March 21.51 UT by Itagati and
Arbour (IAUC 8097) at 14" E and 5" N from the centre of the nearby
spiral galaxy NGC~3169 (Figure \ref{fig_seq}). The SN is projected
on a dust lane, already suggesting that the SN light might be
heavily extinguished. Some days later, \cite{kotak03} and
\cite{matheson03} classified SN~2003cg as a highly-reddened
'normal' Type Ia SN, of epoch a few days before maximum.

NGC~3169 is a nearby ($v_{r}$=1238 \kms), elongated Sa galaxy with
narrow, distorted arms. It has a total B magnitude of 11.08 with a
prominent dust lane. SN~1984E also exploded in NGC~3169. This was
a Type~IIL SN with a B magnitude at maximum of 15.2. Comparison of
the intrinsic colour curve with those of other Type II SNe
\citep{metlova85} yielded \ebv=0.1$\pm$0.05 for this older event.
It exhibited evidence of pre-explosion superwinds
(\citealt{dopita84}; \citealt{gaskell84}; \citealt{gaskell86};
\citealt{henry87}) similar to those of SNe~1994aj, 1996L and
1996al (\citealt{benetti00}).\\

In view of the favorable SN position on the sky
($\alpha=10^{h}14^{m}15\psec97, \delta=
+03^{o}28\arcmin02\parcsec50$, 2000.0), the early discovery epoch
and the proximity of the host galaxy ($v_{r}$=1238 \kms), an
extensive observational campaign was immediately triggered with
all the telescopes available to the ESC. Using 13 different
instrumental configurations, we obtained optical and NIR data from
day -8.5 to +414 relative to B band maximum light. The
observational programme comprised 74 photometric epochs and 35
spectra. \\

The SN~2003cg observations were processed using IRAF\footnote{IRAF
is distributed by the National Optical Astronomy Observatories,
which are operated by the Association of Universities for Research
in Astronomy, Inc, under contract with the National Science
Foundation.} and FIGARO\footnote{http://www.aao.gov.au/figaro/}
routines. For photometry we used a collection of tasks developed
in the IRAF environment by the Padova-Asiago SN Group.

\subsection{Photometric observations} \label{reduc_ph}
Standard pre-processing (trimming, overscan and bias corrections,
flat-fielding) was first carried out. Then, for the early-time
optical photometry, the instrumental optical magnitudes of the SN
and the local standards were measured using the IRAF
point-spread-function (PSF) fitting routine. Specifically, after
fitting and subtracting the background galaxy contamination, we
used Daophot to fit the SN profile with a PSF obtained from a set
of local, unsaturated stars. These local standards in the SN field
(see Figure \ref{fig_seq}) were also used to calibrate the SN
brightness on non-photometric nights. To calibrate the local
standards, and to find the colour terms for the various
instrumental setups, average values for 20 photometric nights were
obtained (marked in Table \ref{tabla_ori_ph_opt}). These were then
calibrated via standard Landolt fields \citep{land92}. Magnitudes
and estimated errors for the local sequence stars are
given in Table \ref{tabla_seq}. \\

For the early-time IR magnitudes, additional reduction steps were
required viz. sky substraction, and image coaddition to improve
the signal-to-noise (see Table \ref{tabla_ori_ph_ir}). As for the
optical photometry, the IR magnitudes were obtained using
PSF-fitting. For night-to-night calibration we used a smaller
local sequence. However, as the number of IR standard fields
observed each night was small, we adopted average colour terms
provided by the telescope teams. The IR magnitudes of the local
sequence stars are given in Table \ref{tabla_seq}.\\

SN~2003cg was also observed at late phases. In this case,
determination of the SN magnitudes made use of template
subtraction. The procedure makes use of the ISIS template
subtraction program \citep{alard00} and runs in the IRAF
environment. A reference image of the host galaxy is obtained long
after the SN image. The two images are then registered
geometrically and photometrically. The better-seeing image is
degraded to that of the poorer one. Finally the reference
(template) image is subtracted from the SN image. The template
images were taken with VLT-UT2+FORS1 on February 1, 2005, around
673 days post-explosion, when the supernova was no longer visible.
The technique was first tested for early phase images when the SN
was bright. It was found that the PSF-fitting and
template-subtraction methods produced results in excellent
agreement even with a variety of instrumental configurations. The
late-time magnitudes of SN~2003cg are shown in Table
\ref{tabla_ori_ph_opt}.\\

Uncertainties in the instrumental photometry were estimated by placing
artificial stars with the same magnitude and profile as the SN, at
positions close (within a few arcsec) to that of the SN, and then
computing the deviations of the artificial star magnitudes. For the
calibration error we adopted the r.m.s. of the observed magnitudes of
the local sequence stars obtained during photometric nights only.  \\

\subsection{Spectroscopic observations} \label{reduc_spec}
The spectra were reduced using standard IRAF (used for optical and
IR reduction) or FIGARO (used for IR reduction only) routines.
This includes bias, overscan (only for the optical spectra) and
flat field correction. We then fitted a polynomial function to the
background flux including that of the galaxy, and removed the
interpolated background by subtraction. Extraction from the
bi-dimensional spectra made use of an optimised extraction
algorithm. The one-dimensional spectra were wavelength calibrated
by comparison with arc-lamp spectra, and flux calibrated using
spectrophotometric standard stars. For most spectra the errors in
the wavelength calibration are $<\pm$2 \AA.  The standard star
spectra were also used to model and remove the telluric
absorptions. The absolute flux calibration of the spectra was
checked against photometry and when necessary, the flux scale was
adjusted. After that, the calibration is typically accurate within
$\pm$10-20\%. The spectroscopic observations are summarised in
Tables \ref{tabla_spec_opt} $\&$ \ref{tabla_spec_ir}.  The tables
give the observation date, the Modified Julian Day, the phase
relative to $B_{max}$, the wavelength
range and the instruments used.\\

\begin{table*}
\centering
\caption{Original optical photometry of SN 2003cg.}
\label{tabla_ori_ph_opt}
\begin{tabular}{lcrcccccc}
\hline
   date  &   MJD   & Phase*&         U        &         B        &         V        &         R        &         I        & Instr.\\
         &          &(days)&                  &                  &                  &                  &                  &       \\
\hline
13/03/03 & 52711.30 & -18.1 &        -         &        -         &        -         & 19.0 &        -         & UNF \\
22/03/03 & 52720.30 & -9.1 &        -         &        -         &        -         & 14.9 &        -         & UNF \\
23/03/03$\triangleleft$ & 52721.94 & -7.5 & $17.03\pm0.19$ & $16.49\pm0.06$ & $15.27\pm0.04$ & $14.58\pm0.02$ & $14.14\pm0.07$ &  CAF \\
24/03/03$\triangleleft$ & 52722.54 & -6.9 & $16.91\pm0.08$ & $16.35\pm0.03$ & $15.24\pm0.03$ & $14.55\pm0.02$ & $14.07\pm0.02$ &  SSO \\
24/03/03 & 52722.56 & -6.9 &        -         & $16.34\pm0.03$ & $15.21\pm0.04$ & $14.50\pm0.03$ & $14.00\pm0.03$ &  BAO \\
25/03/03$\triangleleft$ & 52723.44 & -6.0 & $16.77\pm0.04$ & $16.22\pm0.01$ & $15.09\pm0.01$ & $14.44\pm0.02$ & $13.95\pm0.01$ &  SSO \\
26/03/03 & 52724.53 & -4.9 &        -         & $16.13\pm0.20$ & $14.99\pm0.36$ & $14.29\pm0.06$ &         -        &  BAO \\
26/03/03 & 52724.82 & -4.7 &        -         & $16.09\pm0.02$ & $14.92\pm0.03$ & $14.29\pm0.02$ & $13.85\pm0.03$ &  ASI \\
27/03/03 & 52725.85 & -3.6 &        -         & $16.03\pm0.02$ & $14.84\pm0.03$ & $14.27\pm0.02$ & $13.84\pm0.04$ &  ASI \\
28/03/03$\triangleleft$ & 52726.02 & -3.5 & $16.67\pm0.02$ & $16.08\pm0.03$ & $14.85\pm0.02$ & $14.24\pm0.03$ & $13.77\pm0.02$ &  WFI \\
28/03/03 & 52726.51 & -3.0 &        -         & $16.02\pm0.19$ & $14.82\pm0.03$ & $14.19\pm0.04$ & $13.82\pm0.03$ &  BAO \\
29/03/03 & 52727.02 & -2.5 &        -         & $15.99\pm0.04$ & $14.78\pm0.06$ & $14.18\pm0.03$ & $13.84\pm0.03$ &  ASI \\
29/03/03 & 52727.57 & -1.9 &        -         & $15.98\pm0.06$ & $14.82\pm0.02$ & $14.17\pm0.05$ & $13.80\pm0.05$ &  BAO \\
01/04/03 & 52730.83 &  1.4 &        -         & $15.97\pm0.03$ & $14.72\pm0.05$ & $14.15\pm0.05$ & $13.86\pm0.03$ &  ASI \\
02/04/03$\triangleleft$ & 52731.16 &  1.7 & $16.79\pm0.03$ & $16.05\pm0.02$ & $14.74\pm0.06$ & $14.15\pm0.02$ & $13.88\pm0.03$ &  WFI \\
04/04/03$\triangleleft$ & 52733.12 &  3.7 & $16.87\pm0.09$ & $16.10\pm0.02$ & $14.76\pm0.03$ & $14.18\pm0.02$ & $13.95\pm0.04$ &  WFI \\
04/04/03 & 52733.62 &  4.2 &        -         & $16.04\pm0.05$ & $14.79\pm0.05$ & $14.19\pm0.04$ & $13.97\pm0.03$ &  BAO \\
04/04/03 & 52733.89 &  4.4 & $16.89\pm0.10$ & $16.05\pm0.03$ & $14.83\pm0.03$ & $14.24\pm0.03$ & $14.00\pm0.06$ &  TGD \\
07/04/03 & 52736.51 &  7.1 &        -         & $16.22\pm0.08$ & $14.93\pm0.04$ & $14.35\pm0.05$ & $14.12\pm0.03$ &  BAO \\
07/04/03 & 52736.80 &  7.3 &        -         & $16.25\pm0.04$ & $14.92\pm0.04$ & $14.39\pm0.07$ & $14.08\pm0.05$ &  ASI \\
09/04/03$\triangleleft$ & 52738.90 &  9.4 & $17.26\pm0.02$ & $16.45\pm0.01$ & $14.99\pm0.03$ & $14.56\pm0.02$ & $14.27\pm0.04$ &  TGD \\
11/04/03 & 52740.87 & 11.4 & $17.51\pm0.01$ & $16.65\pm0.05$ & $15.13\pm0.03$ & $14.68\pm0.04$ & $14.37\pm0.03$ &  TGD \\
12/04/03$\triangleleft$ & 52741.90 & 12.4 & $17.69\pm0.04$ & $16.79\pm0.01$ & $15.17\pm0.02$ & $14.73\pm0.04$ & $14.33\pm0.06$ &  CAF \\
14/04/03$\triangleleft$ & 52743.07 & 13.6 & $18.06\pm0.13$ & $16.93\pm0.01$ & $15.19\pm0.02$ & $14.78\pm0.01$ & $14.36\pm0.02$ &  WFI \\
15/04/03 & 52744.02 & 14.6 & $18.10\pm0.03$ & $17.13\pm0.01$ & $15.26\pm0.02$ & $14.81\pm0.01$ & $14.37\pm0.02$ &  WFI \\
16/04/03 & 52745.08 & 15.6 & $18.40\pm0.03$ & $17.22\pm0.04$ & $15.30\pm0.03$ & $14.82\pm0.02$ & $14.30\pm0.03$ &  WFI \\
17/04/03 & 52746.09 & 16.6 & $18.78\pm0.05$ & $17.34\pm0.03$ & $15.33\pm0.02$ & $14.83\pm0.02$ & $14.24\pm0.02$ &  WFI \\
18/04/03 & 52747.11 & 17.6 & $18.93\pm0.04$ & $17.42\pm0.02$ & $15.41\pm0.02$ & $14.84\pm0.01$ & $14.24\pm0.02$ &  WFI \\
19/04/03 & 52748.04 & 18.6 & $19.04\pm0.02$ & $17.57\pm0.02$ & $15.42\pm0.03$ & $14.85\pm0.02$ & $14.22\pm0.02$ &  WFI \\
20/04/03$\triangleleft$ & 52749.07 & 19.6 & $19.26\pm0.05$ & $17.68\pm0.03$ & $15.47\pm0.01$ & $14.85\pm0.01$ & $14.20\pm0.02$ &  WFI \\
21/04/03$\triangleleft$ & 52750.14 & 20.7 & $19.29\pm0.02$ & $17.77\pm0.02$ & $15.53\pm0.02$ & $14.87\pm0.02$ & $14.17\pm0.01$ &  WFI \\
21/04/03 & 52750.98 & 21.5 & $19.30\pm0.03$ & $17.92\pm0.02$ & $15.54\pm0.02$ & $14.90\pm0.02$ & $14.17\pm0.01$ &  WFI \\
23/04/03$\triangleleft$  & 52752.01 & 22.5 & $19.48\pm0.02$ & $17.98\pm0.01$ & $15.60\pm0.01$ & $14.90\pm0.01$ & $14.15\pm0.01$ &  WFI \\
23/04/03 & 52752.95 & 23.5 &        -         & $18.09\pm0.08$ &        -         & $14.93\pm0.02$ & $14.16\pm0.03$ &  ASI \\
24/04/03$\triangleleft$ & 52753.19 & 23.7 & $19.57\pm0.09$ & $18.17\pm0.02$ & $15.65\pm0.03$ & $14.92\pm0.01$ & $14.11\pm0.02$ &  WFI \\
25/04/03$\triangleleft$ & 52754.01 & 24.5 & $19.73\pm0.03$ & $18.26\pm0.01$ & $15.69\pm0.08$ & $14.93\pm0.01$ & $14.15\pm0.01$ &  WFI \\
25/04/03$\triangleleft$ & 52755.00 & 25.5 & $19.84\pm0.02$ & $18.34\pm0.01$ & $15.78\pm0.01$ & $14.99\pm0.02$ & $14.13\pm0.01$ &  WFI \\
26/04/03 & 52755.85 & 26.4 &        -         &        -         & $15.90\pm0.02$ & $15.02\pm0.02$ &        -         &  CAF \\
27/04/03$\triangleleft$ & 52756.11 & 26.6 & $19.91\pm0.02$ & $18.46\pm0.01$ & $15.82\pm0.01$ & $15.06\pm0.01$ & $14.12\pm0.01$ &  WFI \\
28/04/03$\triangleleft$ & 52757.00 & 27.5 & $19.90\pm0.02$ & $18.57\pm0.01$ & $15.94\pm0.01$ & $15.09\pm0.01$ & $14.16\pm0.01$ &  WFI \\
29/04/03 & 52758.51 & 29.1 &        -         &        -         &        -         & $15.18\pm0.03$ & $14.23\pm0.04$ &  BAO \\
30/04/03 & 52759.13 & 29.7 & $20.24\pm0.04$ & $18.80\pm0.01$ & $16.08\pm0.01$ & $15.23\pm0.01$ & $14.25\pm0.02$ &  WFI \\
30/04/03 & 52759.50 & 30.0 &        -         &        -         & $16.12\pm0.11$ & $15.23\pm0.04$ & $14.30\pm0.09$ &  BAO \\
02/05/03 & 52761.15 & 31.7 & $20.28\pm0.02$ & $18.94\pm0.03$ & $16.14\pm0.05$ & $15.38\pm0.02$ & $14.44\pm0.02$ &  WFI \\
02/05/03 & 52761.99 & 32.5 & $20.31\pm0.03$ & $18.98\pm0.05$ & $16.19\pm0.02$ & $15.44\pm0.02$ & $14.52\pm0.02$ &  WFI \\
04/05/03 & 52763.01 & 33.5 & $20.48\pm0.05$ & $19.06\pm0.03$ & $16.25\pm0.02$ & $15.52\pm0.02$ & $14.58\pm0.01$ &  WFI \\
05/05/03 & 52764.99 & 35.5 & $20.43\pm0.03$ & $19.05\pm0.02$ & $16.41\pm0.01$ & $15.63\pm0.03$ & $14.70\pm0.01$ &  WFI \\
07/05/03$\triangleleft$ & 52766.01 & 36.5 & $20.43\pm0.03$ & $19.04\pm0.01$ & $16.44\pm0.01$ & $15.69\pm0.02$ & $14.80\pm0.01$ &  WFI \\
07/05/03 & 52766.85 & 37.4 &        -         &        -         & $16.47\pm0.07$ &         -        &         -        &  ASI \\
07/05/03 & 52766.99 & 37.5 & $20.55\pm0.05$ & $19.10\pm0.02$ & $16.50\pm0.01$ & $15.73\pm0.02$ & $14.87\pm0.02$ &  WFI \\
08/05/03$\triangleleft$ & 52767.85 & 38.4 &        -         & $19.10\pm0.09$ &        -         & $15.79\pm0.03$ & $14.91\pm0.05$ &  ASI \\
09/05/03 & 52768.37 & 38.9 &        -         & $19.05\pm0.07$ & $16.66\pm0.03$ & $15.80\pm0.05$ & $14.96\pm0.01$ &  SSO \\
17/05/03 & 52776.52 & 47.1 &        -         &        -         & $16.99\pm0.04$ & $16.15\pm0.03$ & $15.29\pm0.05$ &  BAO \\
22/05/03 & 52781.94 & 52.5 & $20.64\pm0.07$ & $19.26\pm0.03$ & $17.03\pm0.04$ & $16.37\pm0.04$ & $15.66\pm0.04$ &  TGD \\
23/05/03 & 52782.84 & 53.4 &        -         &        -         & $17.03\pm0.10$ & $16.39\pm0.05$ & $15.69\pm0.03$ &  ASI \\
31/05/03 & 52790.92 & 61.5 & $20.78\pm0.09$ & $19.33\pm0.02$ & $17.39\pm0.04$ & $16.60\pm0.02$ & $16.14\pm0.08$ &  TGO \\
07/06/03 & 52797.36 & 67.9 &        -         &        -         &        -         &        -         & $16.22\pm0.12$ &  SSO \\
19/06/03$\triangleleft$ & 52809.02 & 79.6 &        -         & $19.58\pm0.03$ & $17.75\pm0.01$ & $17.21\pm0.01$ & $16.76\pm0.02$ &  WFI \\
25/06/03 & 52815.01 & 85.5 &        -         & $19.60\pm0.02$ & $17.95\pm0.02$ & $17.38\pm0.03$ & $16.91\pm0.07$ &  WFI \\
\end{tabular}
\end{table*}

\begin{table*}
\centering
\contcaption{}
\begin{tabular}{lcrcccccc}
\hline
   date  &   MJD   & Phase*&         U        &         B        &         V        &         R        &         I        & Instr.\\
         &          &(days)&                  &                  &                  &                  &                  &       \\
\hline
20/11/03 & 52963.20 & 233.7 &       -         &        -         & $21.09\pm0.09$ & $20.75\pm0.06$ &        -         & ASI \\
31/01/04 & 53035.23 & 305.8 &       -         & $23.02\pm0.18$ & $21.68\pm0.04$ & $21.23\pm0.28$ & $20.01\pm0.17$ & TGD \\
19/04/04 & 53114.08 & 384.6 &       -         & $24.12\pm0.07$ & $22.95\pm0.12$ & $22.59\pm0.08$ & $21.58\pm0.02$ & FOR \\
07/05/04 & 53132.97 & 413.5 &       -         & $24.74\pm0.17$ & $22.92\pm0.14$ & $22.22\pm0.05$ & $21.89\pm0.04$ & FOR \\
14/12/04 & 53353.29 & 623.8 &      -         & $>23.60$ & $>22.63$ & $>23.18$ & $>21.53$ & WFI \\
\hline
\end{tabular}
\begin{flushleft}
* Relative to B$_{max}$ (MJD=52729.40)\\
$\triangleleft$ Photometric night\\
UNF = Unfiltered CCD frames taken with a 0.60-m, Japan
(\citep{itagati03}; CAF = Calar Alto 2.2m + CAFOS + CCD SITe
0.53\arcsec/px; SSO = Siding Spring Observatory 2.3m + CCD
0.59\arcsec/px; BAO = Beijing Astronomical Observatory 0.85m + CCD
0.45\arcsec/px; ASI = Asiago 1.82m Copernico telescope + AFOSC
0.47\arcsec/px; TGD = Telescopio Nazionale Galileo + DOLORES
0.28\arcsec/px; TGO = Telescopio Nazionale Galileo + OIG
0.07\arcsec/px;
WFI = ESO/MPI 2.2m + WFI 0.24\arcsec/px; FOR = ESO VLT-UT2 + FORS1 0.20\arcsec/px\\
\end{flushleft}
\end{table*}

\begin{table*}
\centering
\caption{Magnitudes for the local sequence stars identified in the
field of SN 2003cg coded as in Figure \ref{fig_seq}.}
\label{tabla_seq}
\setlength\tabcolsep{2pt} 
\begin{tabular}{ccccccccc}
\hline
~~star~~ & U & B & V & R & I & J & H & K \\
\hline
1        &$17.15\pm0.01$ &$16.04\pm0.01$ &$14.97\pm0.01$ &$14.27\pm0.01$ &$13.73\pm0.01$ &$13.18\pm0.03$ &$12.63\pm0.02$ &$12.45\pm0.01$\\
2        &$17.39\pm0.01$ &$17.46\pm0.01$ &$16.85\pm0.01$ &$16.49\pm0.01$ &$16.08\pm0.01$ &       -         &       -         &       -        \\
3        &$18.43\pm0.02$ &$17.32\pm0.01$ &$16.14\pm0.01$ &$15.37\pm0.01$ &$14.69\pm0.01$ &$14.00\pm0.04$ &$13.26\pm0.04$ &$13.11\pm0.03$\\
4        &$16.96\pm0.01$ &$16.67\pm0.01$ &$15.91\pm0.01$ &$15.47\pm0.01$ &$15.03\pm0.01$ &$14.63\pm0.03$ &$14.10\pm0.03$ &$13.98\pm0.04$\\
5        &$19.70\pm0.03$ &$18.55\pm0.01$ &$17.09\pm0.01$ &$16.18\pm0.00$ &$15.27\pm0.02$ &       -         &       -         &       -        \\
6        &$20.54\pm0.07$ &$19.52\pm0.01$ &$18.07\pm0.01$ &$17.18\pm0.02$ &$16.41\pm0.02$ &       -         &       -         &       -        \\
7        &$13.70\pm0.01$ &$13.75\pm0.01$ &$13.25\pm0.01$ &$12.88\pm0.01$ &$12.55\pm0.02$ &       -         &       -         &       -        \\
8        &$17.56\pm0.01$ &$17.29\pm0.01$ &$16.52\pm0.01$ &$16.13\pm0.01$ &$15.69\pm0.01$ &       -         &         -         &       -        \\
9        &$16.41\pm0.01$ &$16.34\pm0.01$ &$15.61\pm0.01$ &$15.19\pm0.01$ &$14.75\pm0.01$ &$14.26\pm0.02$ &$13.77\pm0.01$ &$13.67\pm0.02$\\
10       &$18.22\pm0.01$ &$18.32\pm0.01$ &$17.79\pm0.01$ &$17.48\pm0.01$ &$17.10\pm0.01$ &$16.74\pm0.04$ &$16.48\pm0.02$ &$16.37\pm0.01$\\
11       &       -         &$21.18\pm0.01$ &$19.57\pm0.02$ &$18.24\pm0.01$ &$16.41\pm0.01$ &$15.08\pm0.03$ &$14.48\pm0.03$ &$14.21\pm0.04$\\
12       &$18.08\pm0.01$ &$17.95\pm0.01$ &$17.28\pm0.01$ &$16.90\pm0.01$ &$16.54\pm0.01$ &$16.21\pm0.03$ &$15.85\pm0.01$ &$15.85\pm0.04$\\
13       &$19.73\pm0.01$ &$18.41\pm0.01$ &$17.16\pm0.01$ &$16.37\pm0.01$ &$15.75\pm0.01$ &       -       &       -         &       -        \\
14       &$14.67\pm0.02$ &$14.46\pm0.01$ &$14.27\pm0.01$ &$14.12\pm0.01$ &$13.98\pm0.01$ &       -         &         -         &       -        \\
15       &$18.70\pm0.01$ &$17.42\pm0.01$ &$16.05\pm0.01$ &$15.19\pm0.01$ &$14.32\pm0.01$ &       -         &         -         &       -        \\
\hline
\end{tabular}
\end{table*}

\begin{table*}
\centering
\begin{minipage}{125mm}
\caption{Original near-IR photometry of SN 2003cg.}
\label{tabla_ori_ph_ir}
\begin{tabular}{lcrcccc}
\hline
 Date      &    MJD  &    Phase* &     J    &     H     &     K    &   Instr.\\
           &         &    (days)&           &            &          &        \\
\hline
25/03/03$\triangleleft$ & 52723.06 & -6.4 & $13.71\pm0.04$ & $13.68\pm0.05$ & $13.50\pm0.01$ & SofI \\
28/03/03$\triangleleft$ & 52726.05 & -3.4 & $13.56\pm0.08$ & $13.61\pm0.10$ & $13.42\pm0.01$ & SofI \\
10/04/03$\triangleleft$ & 52739.96 & 10.5 & $15.18\pm0.01$ & $14.03\pm0.01$ & $13.90\pm0.02$ & SofI \\
12/04/03$\triangleleft$ & 52741.98 & 12.5 & $15.24\pm0.12$ & $13.92\pm0.10$ & $13.75\pm0.01$ & SofI \\
20/04/03$\triangleleft$ & 52749.17 & 19.7 & $14.89\pm0.17$ & $13.66\pm0.01$ & $13.59\pm0.01$ & SofI \\
24/04/03$\triangleleft$ & 52753.18 & 23.7 & $14.76\pm0.13$ & $13.61\pm0.03$ & $13.51\pm0.02$ & SofI \\
01/05/03$\triangleleft$ & 52760.03 & 30.6 & $14.49\pm0.01$ & $13.83\pm0.02$ & $13.82\pm0.02$ & SofI \\
09/05/03$\triangleleft$ & 52768.98 & 39.5 & $15.15\pm0.09$ & $14.26\pm0.10$ & $14.34\pm0.09$ & SofI \\
28/05/03$\triangleleft$ & 52787.95 & 58.5 & $16.55\pm0.08$ & $15.08\pm0.01$ & $15.11\pm0.02$ & SofI \\
16/04/04$\triangleleft$ & 53110.67 & 381.2 & $19.80\pm0.81$ & $19.00\pm0.51$ & $20.48\pm0.69$ & Isaac \\
18/05/04$\triangleleft$ & 53143.01 & 413.5 & $20.09\pm0.21$ & $19.29\pm0.13$ & $20.55\pm0.32$ & Isaac \\
\hline
\end{tabular}

\begin{flushleft}
* Relative to B$_{max}$ (MJD=52729.40)\\
$\triangleleft$ Photometric night\\
SofI = ESO NTT + SofI 0.29\arcsec/px; Isaac = ESO VLT-UT1 + Isaac 0.15\arcsec/px\\
\end{flushleft}
\end{minipage}
\end{table*}

\begin{table}
\centering
\caption{Optical spectroscopy of SN 2003cg.}
\label{tabla_spec_opt}
\begin{tabular}{lcrcc}
\hline
 Date    &   MJD    &Phase*&  Range    &Instr.\\
         &          &(days)& ($\AA$)   &      \\
\hline
22/03/03 & 52720.99 & -8.5 & 4227-6981 & WHT \\
23/03/03 & 52721.89 & -7.6 & 3700-9350 & CAF \\
24/03/03 & 52722.56 & -6.9 & 3650-9132 & SSO \\
25/03/03 & 52723.45 & -6.0 & 3720-9153 & SSO \\
26/03/03 & 52724.89 & -4.6 & 3650-9350 & ASI \\
29/03/03 & 52727.06 & -2.4 & 4000-7773 & ASI \\
29/03/03 & 52727.60 & -1.9 & 4000-8500 & BAO \\
30/03/03 & 52728.90 & -0.6 & 3750-9196 & CAF \\
01/04/03 & 52730.90 &  1.4 & 3900-9350 & ASI \\
04/04/03 & 52733.90 &  4.4 & 3500-7968 & TGD \\
07/04/03 & 52736.87 &  7.4 & 3700-7752 & ASI \\
10/04/03 & 52739.08 &  9.6 & 3380-7970 & TGD \\
11/04/03 & 52740.89 & 11.4 & 3400-8050 & TGD \\
12/04/03 & 52741.86 & 12.4 & 3400-9250 & CAF \\
16/04/03 & 52745.90 & 16.4 & 3500-9250 & INT \\
19/04/03 & 52748.88 & 19.4 & 3800-8050 & INT \\
23/04/03 & 52752.91 & 23.4 & 3800-7750 & INT \\
23/04/03 & 52752.92 & 23.5 & 3800-7750 & ASI \\
26/04/03 & 52755.89 & 26.4 & 3800-9350 & CAF \\
28/04/03 & 52758.00 & 28.5 & 4250-7750 & INT \\
08/05/03 & 52767.89 & 38.4 & 4000-7770 & ASI \\
09/05/03 & 52768.37 & 38.9 & 6100-8920 & SSO \\
13/05/03 & 52772.93 & 43.5 & 3810-9250 & INT \\
22/05/03 & 52781.97 & 52.5 & 3500-8050 & TGD \\
19/04/04 & 53114.08 & 384.6& 4000-8050 & FOR \\
\hline
\end{tabular}
\begin{flushleft}
* Relative to B$_{max}$ (MJD=52729.40)\\
WHT = William Herschel Telescope + ISIS; CAF = Calar Alto 2.2m +
CAFOS; SSO = Siding Spring Observatory 2.3m + Double Beam
Spectrograph; ASI = Asiago 1.82m Copernico Telescope + AFOSC; BAO
= Beijing Astronomical Observatory 0.85m + CCD; TGD = Telescopio
Nazionale Galileo +
DOLORES; INT = Isaac Newton Telescope + IDS; FOR = ESO VLT-UT2 + FORS1\\
\end{flushleft}
\end{table}

\begin{table}
\centering
\caption{Infrared spectroscopy of SN 2003cg.}
\label{tabla_spec_ir}
\begin{tabular}{lcrcc}
\hline
 Date    &   MJD    &Phase*&  Range    &Instr.\\
         &          &(days)& ($\AA$)   &      \\
\hline
25/03/03 & 52723.14 & -6.5 & 7970-24984 & SofI+UKI \\
27/03/03 & 52725.35 & -4.1 & 14002-25035 & UKI \\
28/03/03 & 52726.02 & -3.5 & 9367-24980 & SofI \\
04/04/03 & 52733.42 &  3.9 & 14242-25030 & UKI \\
14/04/03 & 52743.40 & 13.9 & 10378-25037 & UKI \\
20/04/03 & 52749.10 & 19.6 & 9405-24990 & SofI \\
21/04/03 & 52750.23 & 20.8 & 9400-25000 & SofI \\
24/04/03 & 52753.28 & 23.8 & 10381-25033 & UKI \\
30/04/03 & 52760.10 & 30.5 & 9398-24990 & SofI \\
10/05/03 & 52769.60 & 40.1 & 9397-24990 & SofI \\
\hline
\end{tabular}
\begin{flushleft}
* Relative to B$_{max}$ (MJD=52729.40)\\
SofI = ESO NTT + SofI; UKI = United Kingdom Infrared Telescope + CGS4\\
\end{flushleft}
\end{table}

\subsection{Correction to standard photometric bands} \label{s_corr}
As indicated in Tables \ref{tabla_ori_ph_opt} $\&$
\ref{tabla_ori_ph_ir}, we used ten different instruments to
collect the photometry of SN 2003cg from U to K band. It is
desirable to convert the photometry of the target to a standard
system. However, this is made difficult by the non-stellar form of
the SN~Ia spectrum. Nevertheless, several authors have shown that,
with care, such a correction can minimise systematic errors
(\citealt{suntzeff00}; \citealt{stritz02}; \citealt{krisc03};
\citealt{pignata04}; \citealt{pignata04b}). The procedure involved
is sometimes called the S-correction (\citealt{stritz02}). We
applied S-corrections to our data following the method of
\cite{pignata04}. The main challenge was to determine the
instrument/filter/detector system efficiency for each observation.
Once the instrumental response functions were constructed, we
calculated an instrumental zero-point for each passband. We then
derived synthetic colour terms for the optical using a set of
spectrophotometric standard stars (\citealt{hamuy92};
\citealt{hamuy94}) and for the IR, using the spectra of Vega,
Sirius and the Sun. With this information we then calculated the
S-corrections for the BVRIJHK bands\footnote{We did not compute an
S-correction for the U~band since very few spectra covered this
wavelength range.} by using the best flux-calibrated spectra of
SN~2003cg. As the epochs of these spectra did not fully match the
early-time photometric coverage, we enhanced the S-correction
spectral database by adding a set of spectra of SN~1994D
(\citealt{patat96}), artificially reddened to match those of
SN~2003cg (see Section \ref{indredd} for more details). The
features of the SNe~1994D and 2003cg spectra are very similar. At
late times (nebular era) we have only one optical spectrum of SN
2003cg. We therefore used the spectra of other normal SNe Ia:
1992A (\citealt{suntzeff96}); 1994D (\citealt{patat96}), 1996X
(\citealt{salvo01}), 2002bo (\citealt{benetti04}), 2002dj (Pignata
et al. in preparation) and 2004eo (Pastorello et al. in
preparation).  For the early-time NIR corrections we used the
extensive spectral sequences of SN~2002bo (\citealt{benetti04})
together with the spectra of SN~2003cg. NIR S-corrections at
nebular epochs were not computed owing to the lack of contemporary
IR spectra.

Once the synthetic magnitudes for each instrumental setup were
obtained, we fitted a low order polynomial to the differences
between these and those of the standard system passbands in order
to correct for the non-stellar nature of the SN spectra. Some
dispersion in the S-corrections relative to a given instrument,
resulted from flux calibration errors in the input observed
spectra plus the fact that we used spectra from different
SNe. Assuming that for each instrument the S-correction changed
smoothly with time, we calculated the r.m.s. deviation with
respect to the polynomial fit and used this as an estimate of the
S-correction errors.\\

\begin{figure}
\psfig{figure=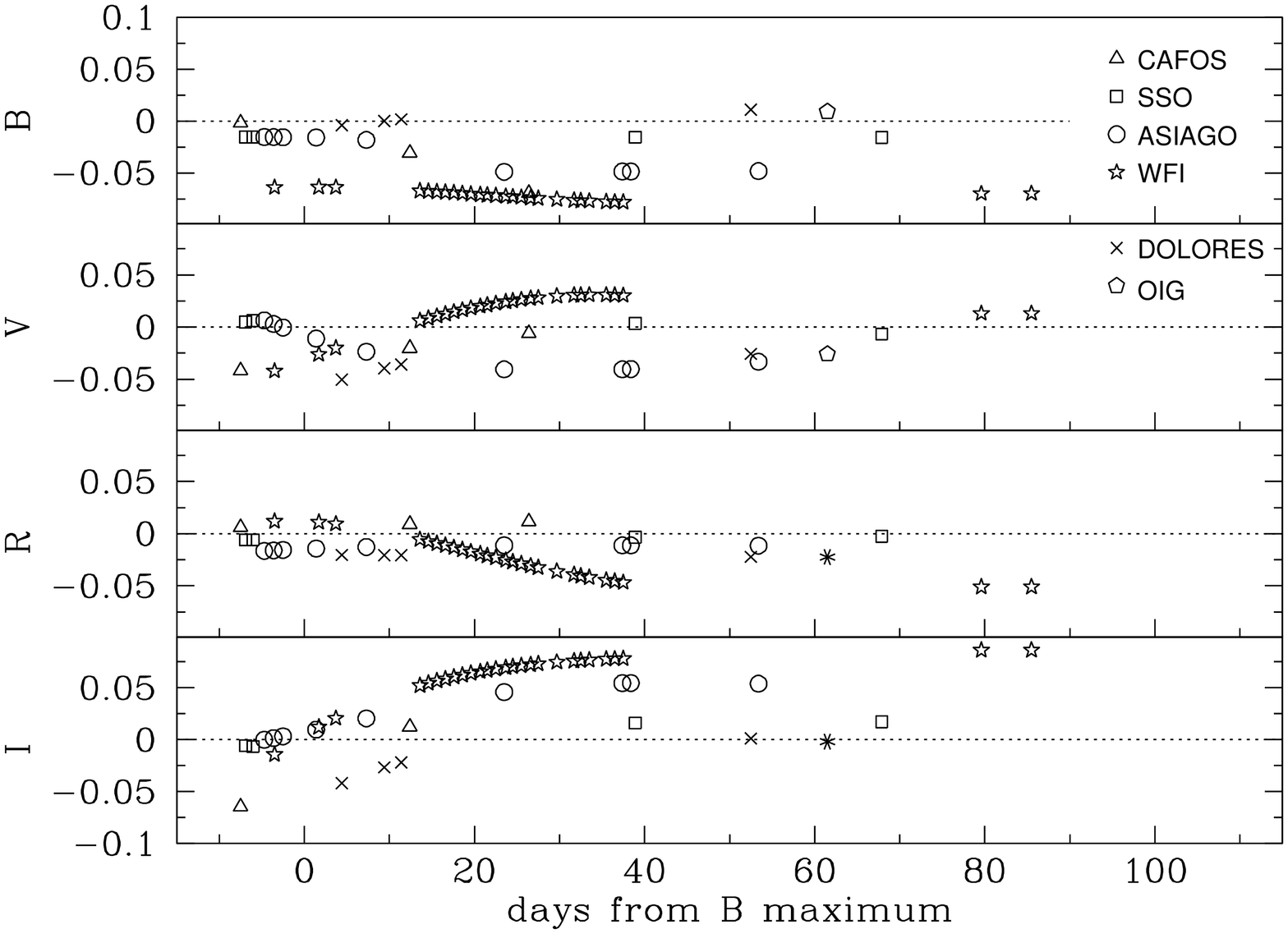,angle=270,width=9.5cm,height=7.2cm}
\psfig{figure=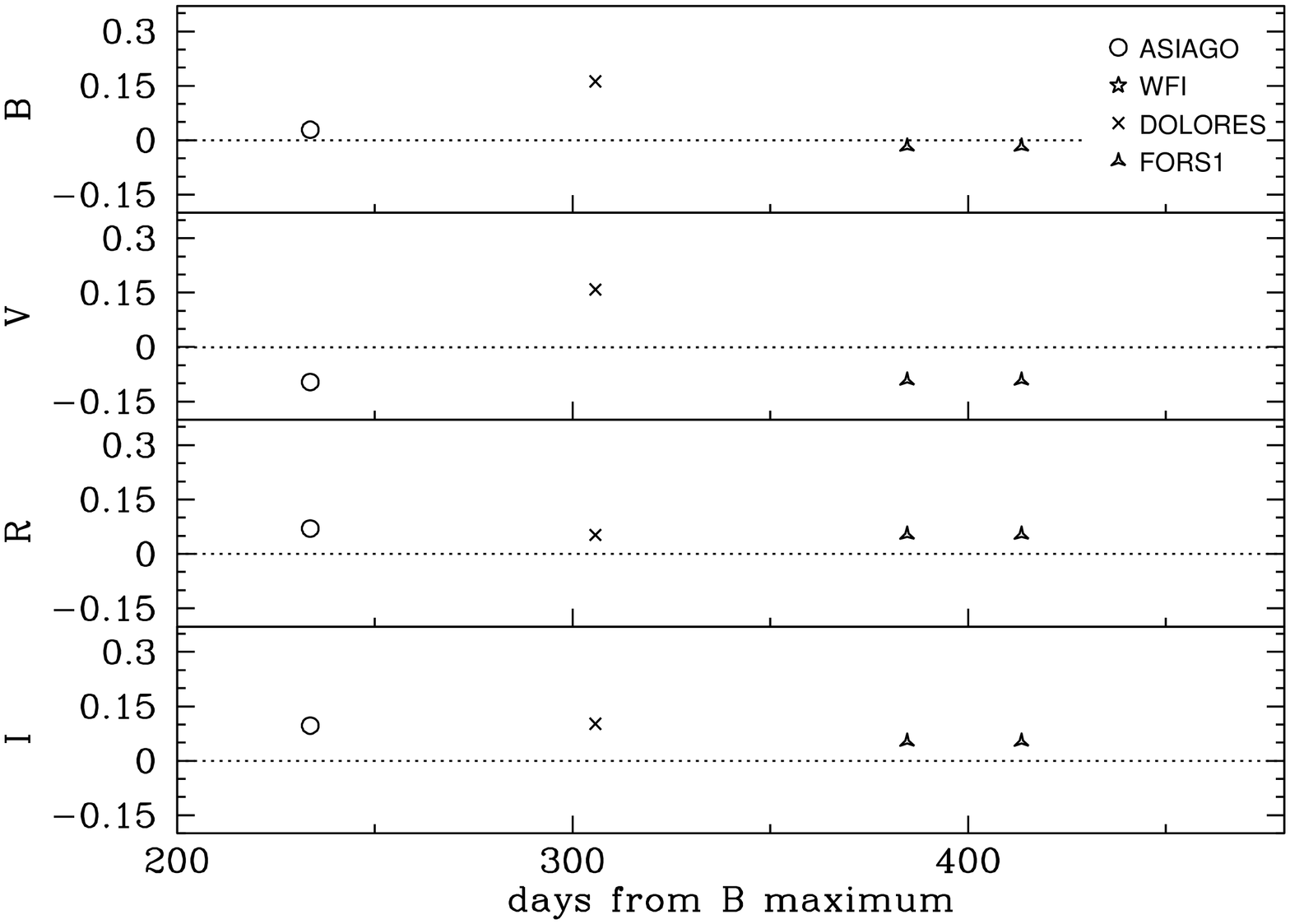,angle=270,width=9.5cm,height=7.2cm}
\psfig{figure=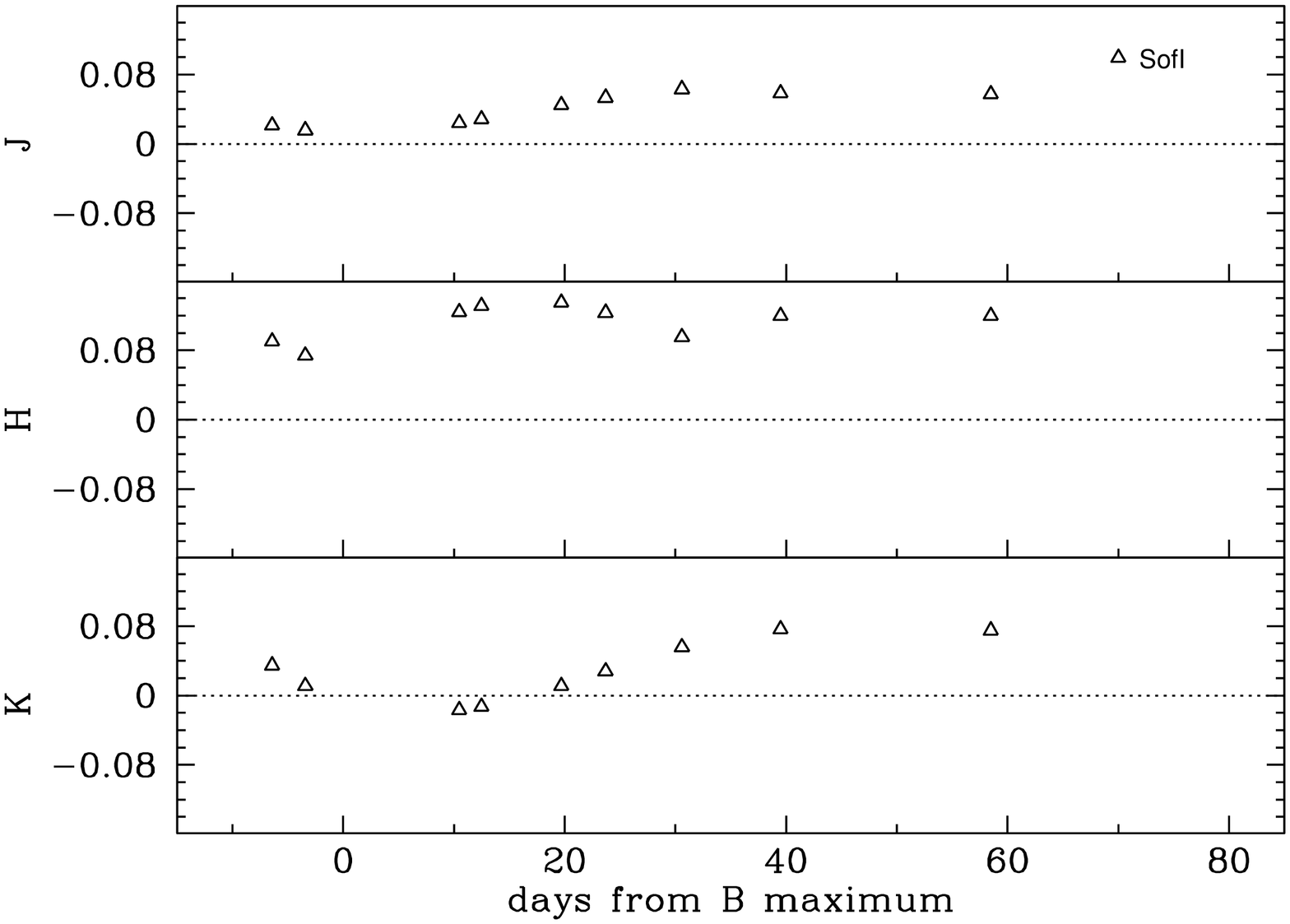,angle=270,width=9.5cm,height=7.2cm}
\caption{Summary of the S-corrections adopted for the BVRI bands
of the different instruments (see legend) at early (top panel) and
late (middle panel) times.  The early-time S-corrections for the
NIR bands are also shown (bottom panel). These corrections should
be added to the first-order corrected SN~2003cg magnitudes in
order to convert to the standard system values. The dotted line
shows the zero correction.} \label{fig_scorr}
\end{figure}

The S-corrections for SN 2003cg in the BVRI bands at early times
is generally small ($\leq0.09$), as seen in Table
\ref{tabla_scorr} and in Figure \ref{fig_scorr} (top). The 0.85m
Beijing telescope data were not S-corrected since it was not
possible to find the basic information needed to calculate the
global response. Nonetheless, the BAO data are in good agreement
with the S-corrected data from other instruments as can be seen in
Figure \ref{fig_lightcurv}. Our S-correction values are comparable
to those obtained by \cite{pignata04} for other SNe~Ia observed
with the same instruments the largest corrections being those in
the I band. As in \cite{pignata04}, the WFI filter passbands
differ from the standard ones by more than do any other
instruments.

The central panel of Figure \ref{fig_scorr} shows the S-correction
estimated for the SN~2003cg BVRI magnitudes from 200 to 450 days
after B~maximum. This work represents one of only a few examples
where the S-correction has been computed at late phases. As
\cite{pignata04b} has already discussed, these corrections are
generally larger than around maximum light, especially in the
B~band. In fact, at late epochs the SNe~Ia emit most of their flux
in a few strong emission lines and so even small shifts in the
filter passbands can produce significant variations in the
instrumental magnitudes. We find evidence of dispersion in the
late-time S-correction due to the low signal-to-noise ratio of the
spectra and to residual galaxy background contamination, but there
are no trends in the corrections for each instrument and therefore
we have adopted an average constant value.

Table \ref{tabla_scorrIR} and Figure \ref{fig_scorr} (bottom)
contain the NIR S-corrections for SofI. We used the same method as
described above to place the magnitudes on the standard system of
\cite{persson98}. As we have mentioned before, to determine the
colour terms we used the synthetic photometry of Vega, Sirius and
the Sun. The corrections were, in general, quite small for
SN~2003cg in JK but increased to about 0.13 magnitudes for the H
band.

In Tables \ref{tabla_ori_ph_opt} $\&$ \ref{tabla_ori_ph_ir} we present
the supernova magnitudes. These have only been first-order corrected
using the colour equation of the instruments.  However, in the rest of
the paper the magnitudes used have been fully S-corrected as
detailed in Tables \ref{tabla_scorr} $\&$ \ref{tabla_scorrIR}.  \\

\begin{table*}
\centering
\caption{Optical S-correction to be added to the data in Table
\ref{tabla_ori_ph_opt}, in order to convert the SN magnitudes to
the Bessell (1990) system.}\label{tabla_scorr}
\begin{tabular}{lcrrrrrc}
\hline
   date   &   MJD    & Phase*&          B         &          V        &         R         &         I         &$Instr.^1$\\
          &          &(days) &                    &                   &                   &                   &               \\
\hline
23/03/03 & 52721.94 & -7.5  & $-0.002(\pm0.010)$ & $-0.041(\pm0.006)$ & $ 0.006(\pm0.001)$ & $-0.065(\pm0.003)$ &  CAF \\
24/03/03 & 52722.54 & -6.9  & $-0.015(\pm0.008)$ & $ 0.005(\pm0.003)$ & $-0.006(\pm0.005)$ & $-0.006(\pm0.002)$ &  SSO \\
25/03/03 & 52723.44 & -6.0  & $-0.015(\pm0.008)$ & $ 0.006(\pm0.003)$ & $-0.006(\pm0.005)$ & $-0.007(\pm0.002)$ &  SSO \\
26/03/03 & 52724.82 & -4.7  & $-0.015(\pm0.004)$ & $ 0.006(\pm0.009)$ & $-0.016(\pm0.005)$ & $ 0.000(\pm0.009)$ &  ASI \\
27/03/03 & 52725.85 & -3.6  & $-0.016(\pm0.004)$ & $ 0.003(\pm0.009)$ & $-0.016(\pm0.005)$ & $ 0.002(\pm0.009)$ &  ASI \\
28/03/03 & 52726.02 & -3.5  & $-0.064(\pm0.010)$ & $-0.042(\pm0.009)$ & $ 0.012(\pm0.008)$ & $-0.014(\pm0.007)$ &  WFI \\
29/03/03 & 52727.02 & -2.5  & $-0.016(\pm0.004)$ & $-0.001(\pm0.009)$ & $-0.016(\pm0.005)$ & $ 0.003(\pm0.009)$ &  ASI \\
01/04/03 & 52730.83 &  1.4  & $-0.016(\pm0.004)$ & $-0.011(\pm0.009)$ & $-0.014(\pm0.005)$ & $ 0.010(\pm0.009)$ &  ASI \\
02/04/03 & 52731.16 &  1.7  & $-0.064(\pm0.010)$ & $-0.026(\pm0.009)$ & $ 0.011(\pm0.008)$ & $ 0.012(\pm0.007)$ &  WFI \\
04/04/03 & 52733.12 &  3.7  & $-0.064(\pm0.010)$ & $-0.020(\pm0.009)$ & $ 0.009(\pm0.008)$ & $ 0.021(\pm0.007)$ &  WFI \\
04/04/03 & 52733.89 &  4.4  & $-0.004(\pm0.006)$ & $-0.050(\pm0.008)$ & $-0.020(\pm0.002)$ & $-0.042(\pm0.005)$ &  TGD \\
07/04/03 & 52736.80 &  7.3  & $-0.018(\pm0.004)$ & $-0.024(\pm0.009)$ & $-0.013(\pm0.005)$ & $ 0.021(\pm0.009)$ &  ASI \\
09/04/03 & 52738.90 &  9.4  & $ 0.001(\pm0.006)$ & $-0.039(\pm0.008)$ & $-0.021(\pm0.002)$ & $-0.027(\pm0.005)$ &  TGD \\
11/04/03 & 52740.87 & 11.4  & $ 0.002(\pm0.006)$ & $-0.036(\pm0.008)$ & $-0.021(\pm0.002)$ & $-0.022(\pm0.005)$ &  TGD \\
12/04/03 & 52741.90 & 12.4  & $-0.031(\pm0.010)$ & $-0.020(\pm0.006)$ & $ 0.009(\pm0.001)$ & $ 0.012(\pm0.003)$ &  CAF \\
14/04/03 & 52743.07 & 13.6  & $-0.067(\pm0.010)$ & $ 0.006(\pm0.009)$ & $-0.005(\pm0.008)$ & $ 0.052(\pm0.007)$ &  WFI \\
15/04/03 & 52744.02 & 14.6  & $-0.068(\pm0.010)$ & $ 0.008(\pm0.009)$ & $-0.007(\pm0.008)$ & $ 0.054(\pm0.007)$ &  WFI \\
16/04/03 & 52745.08 & 15.6  & $-0.068(\pm0.010)$ & $ 0.010(\pm0.009)$ & $-0.009(\pm0.008)$ & $ 0.057(\pm0.007)$ &  WFI \\
17/04/03 & 52746.09 & 16.6  & $-0.069(\pm0.010)$ & $ 0.013(\pm0.009)$ & $-0.011(\pm0.008)$ & $ 0.059(\pm0.007)$ &  WFI \\
18/04/03 & 52747.11 & 17.6  & $-0.069(\pm0.010)$ & $ 0.014(\pm0.009)$ & $-0.013(\pm0.008)$ & $ 0.060(\pm0.007)$ &  WFI \\
19/04/03 & 52748.04 & 18.6  & $-0.070(\pm0.010)$ & $ 0.016(\pm0.009)$ & $-0.015(\pm0.008)$ & $ 0.062(\pm0.007)$ &  WFI \\
20/04/03 & 52749.07 & 19.6  & $-0.070(\pm0.010)$ & $ 0.018(\pm0.009)$ & $-0.017(\pm0.008)$ & $ 0.064(\pm0.007)$ &  WFI \\
21/04/03 & 52750.14 & 20.7  & $-0.071(\pm0.010)$ & $ 0.020(\pm0.009)$ & $-0.019(\pm0.008)$ & $ 0.066(\pm0.007)$ &  WFI \\
21/04/03 & 52750.98 & 21.5  & $-0.071(\pm0.010)$ & $ 0.021(\pm0.009)$ & $-0.021(\pm0.008)$ & $ 0.067(\pm0.007)$ &  WFI \\
23/04/03 & 52752.01 & 22.5  & $-0.072(\pm0.010)$ & $ 0.023(\pm0.009)$ & $-0.023(\pm0.008)$ & $ 0.068(\pm0.007)$ &  WFI \\
23/04/03 & 52752.95 & 23.5  & $-0.049(\pm0.004)$ & $-0.041(\pm0.009)$ & $-0.011(\pm0.005)$ & $ 0.046(\pm0.009)$ &  ASI \\
24/04/03 & 52753.19 & 23.7  & $-0.072(\pm0.010)$ & $ 0.024(\pm0.009)$ & $-0.025(\pm0.008)$ & $ 0.069(\pm0.007)$ &  WFI \\
25/04/03 & 52754.01 & 24.5  & $-0.073(\pm0.010)$ & $ 0.025(\pm0.009)$ & $-0.027(\pm0.008)$ & $ 0.070(\pm0.007)$ &  WFI \\
25/04/03 & 52755.00 & 25.5  & $-0.073(\pm0.010)$ & $ 0.026(\pm0.009)$ & $-0.029(\pm0.008)$ & $ 0.071(\pm0.007)$ &  WFI \\
26/04/04 & 52755.85 & 26.4  & $-0.069(\pm0.010)$ & $-0.006(\pm0.006)$ & $ 0.012(\pm0.001)$ & $-0.108(\pm0.003)$ &  CAF \\
27/04/03 & 52756.11 & 26.6  & $-0.074(\pm0.010)$ & $ 0.027(\pm0.009)$ & $-0.031(\pm0.008)$ & $ 0.072(\pm0.007)$ &  WFI \\
28/04/03 & 52757.00 & 27.5  & $-0.074(\pm0.010)$ & $ 0.028(\pm0.009)$ & $-0.032(\pm0.008)$ & $ 0.073(\pm0.007)$ &  WFI \\
30/04/03 & 52759.13 & 29.7  & $-0.075(\pm0.010)$ & $ 0.030(\pm0.009)$ & $-0.036(\pm0.008)$ & $ 0.075(\pm0.007)$ &  WFI \\
02/05/03 & 52761.15 & 31.7  & $-0.076(\pm0.010)$ & $ 0.030(\pm0.009)$ & $-0.039(\pm0.008)$ & $ 0.076(\pm0.007)$ &  WFI \\
02/05/03 & 52761.99 & 32.5  & $-0.077(\pm0.010)$ & $ 0.031(\pm0.009)$ & $-0.040(\pm0.008)$ & $ 0.076(\pm0.007)$ &  WFI \\
04/05/03 & 52763.01 & 33.5  & $-0.077(\pm0.010)$ & $ 0.031(\pm0.009)$ & $-0.042(\pm0.008)$ & $ 0.077(\pm0.007)$ &  WFI \\
05/05/03 & 52764.99 & 35.5  & $-0.078(\pm0.010)$ & $ 0.031(\pm0.009)$ & $-0.044(\pm0.008)$ & $ 0.078(\pm0.007)$ &  WFI \\
07/05/03 & 52766.01 & 36.5  & $-0.078(\pm0.010)$ & $ 0.031(\pm0.009)$ & $-0.046(\pm0.008)$ & $ 0.078(\pm0.007)$ &  WFI \\
07/05/03 & 52766.85 & 37.4  & $-0.049(\pm0.004)$ & $-0.040(\pm0.009)$ & $-0.011(\pm0.005)$ & $ 0.054(\pm0.009)$ &  ASI \\
07/05/03 & 52766.99 & 37.5  & $-0.078(\pm0.010)$ & $ 0.030(\pm0.009)$ & $-0.047(\pm0.008)$ & $ 0.078(\pm0.007)$ &  WFI \\
08/05/03 & 52767.85 & 38.4  & $-0.049(\pm0.004)$ & $-0.040(\pm0.009)$ & $-0.011(\pm0.005)$ & $ 0.055(\pm0.009)$ &  ASI \\
09/05/03 & 52768.37 & 38.9  & $-0.016(\pm0.008)$ & $ 0.003(\pm0.003)$ & $-0.003(\pm0.005)$ & $ 0.016(\pm0.002)$ &  SSO \\
22/05/03 & 52781.94 & 52.5  & $ 0.011(\pm0.006)$ & $-0.026(\pm0.008)$ & $-0.022(\pm0.002)$ & $ 0.001(\pm0.005)$ &  TGD \\
23/05/03 & 52782.84 & 53.4  & $-0.048(\pm0.004)$ & $-0.033(\pm0.009)$ & $-0.011(\pm0.005)$ & $ 0.054(\pm0.009)$ &  ASI \\
31/05/03 & 52790.92 & 61.5  & $ 0.009(\pm0.006)$ & $-0.026(\pm0.008)$ & $-0.022(\pm0.002)$ & $-0.002(\pm0.005)$ &  TGO \\
07/06/03 & 52797.36 & 67.9  & $-0.016(\pm0.008)$ & $-0.007(\pm0.003)$ & $-0.002(\pm0.005)$ & $ 0.017(\pm0.002)$ &  SSO \\
19/06/03 & 52809.02 & 79.6  & $-0.070(\pm0.010)$ & $ 0.013(\pm0.009)$ & $-0.051(\pm0.008)$ & $ 0.086(\pm0.007)$ &  WFI \\
25/06/03 & 52815.01 & 85.5  & $-0.070(\pm0.010)$ & $ 0.013(\pm0.009)$ & $-0.051(\pm0.008)$ & $ 0.086(\pm0.007)$ &  WFI \\
20/11/03 & 52963.20 & 233.7 & $ 0.029(\pm0.008)$ & $-0.096(\pm0.021)$ & $ 0.070(\pm0.028)$ & $ 0.097(\pm0.097)$ &  ASI \\
31/01/04 & 53035.23 & 305.8 & $ 0.162(\pm0.027)$ & $ 0.158(\pm0.051)$ & $ 0.052(\pm0.021)$ & $ 0.102(\pm0.023)$ &  TGD \\
19/04/04 & 53114.08 & 384.6 & $-0.017(\pm0.009)$ & $-0.092(\pm0.022)$ & $ 0.073(\pm0.024)$ & $ 0.053(\pm0.019)$ &  FOR \\
07/05/04 & 53132.97 & 413.5 & $-0.017(\pm0.009)$ & $-0.092(\pm0.022)$ & $ 0.073(\pm0.024)$ & $ 0.053(\pm0.019)$ &  FOR \\
\hline
\end{tabular}
\begin{flushleft}
* Relative to B$_{max}$ (MJD=52729.40)\\
$^1$ See note to Table \ref{tabla_ori_ph_opt} for the telescope coding.\\
\end{flushleft}
\end{table*}

\begin{table*}
\centering
\caption{NIR S-corrections to be added to the data in Table
\ref{tabla_ori_ph_ir} in order to convert the SN magnitudes to the
Persson et al. (1998) system.}\label{tabla_scorrIR}
\begin{tabular}{lcrrrrc}
\hline
   date   &   MJD    & Phase*&          J         &          H        &         K         &$Instr.^1$\\
          &          &(days) &                    &                   &                   &          \\
\hline
25/03/03 & 52723.56 & -6.4 & $ 0.021(\pm0.017)$ & $ 0.090(\pm0.023)$ & $ 0.035(\pm0.019)$ & SofI \\
28/03/03 & 52726.55 & -3.4 & $ 0.016(\pm0.017)$ & $ 0.074(\pm0.023)$ & $ 0.011(\pm0.019)$ & SofI \\
10/04/03 & 52740.46 & 10.5 & $ 0.024(\pm0.017)$ & $ 0.124(\pm0.023)$ & $-0.017(\pm0.019)$ & SofI \\
12/04/03 & 52742.48 & 12.5 & $ 0.028(\pm0.017)$ & $ 0.131(\pm0.023)$ & $-0.013(\pm0.019)$ & SofI \\
20/04/03 & 52749.67 & 19.7 & $ 0.045(\pm0.017)$ & $ 0.135(\pm0.023)$ & $ 0.011(\pm0.019)$ & SofI \\
24/04/03 & 52753.68 & 23.7 & $ 0.053(\pm0.017)$ & $ 0.123(\pm0.023)$ & $ 0.028(\pm0.019)$ & SofI \\
01/05/03 & 52760.53 & 30.6 & $ 0.063(\pm0.017)$ & $ 0.095(\pm0.023)$ & $ 0.056(\pm0.019)$ & SofI \\
09/05/03 & 52769.48 & 39.5 & $ 0.058(\pm0.017)$ & $ 0.120(\pm0.023)$ & $ 0.077(\pm0.019)$ & SofI \\
28/05/03 & 52788.45 & 58.5 & $ 0.057(\pm0.017)$ & $ 0.120(\pm0.023)$ & $ 0.075(\pm0.019)$ & SofI \\
\hline
\end{tabular}
\begin{flushleft}
* Relative to B$_{max}$ (MJD=2452729.40)\\
$^1$ See note to Table \ref{tabla_ori_ph_ir} for the telescope coding.\\
\end{flushleft}
\end{table*}

\section{The reddening problem} \label{indredd}
\begin{figure}
\psfig{figure=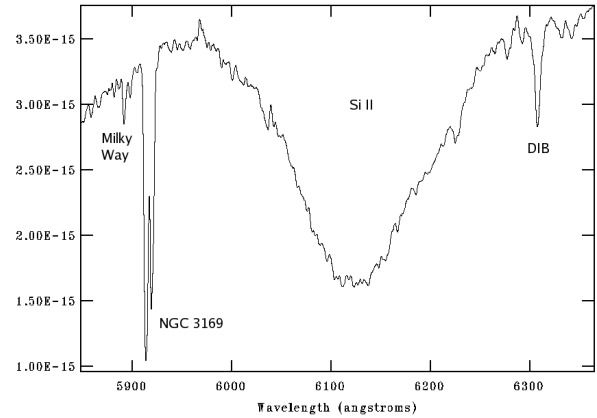,width=8.8cm,height=6.38cm}
\caption{Detail of the interstellar \NaI D region in the
classification spectrum of SN~2004cg obtained on 2003 March 23
with the William Herschel Telescope (+ISIS). The components due to
the Galaxy and NGC~3169 are clearly distinguishable. Also visible
is a narrow absorption of a DIB at $\sim$ 6282 $\AA$.}
\label{fig_naid}
\end{figure}

SN~2003cg appears projected into a dust lane of NGC 3169 (Figure
\ref{fig_seq}) and there are a number of indications that
reddening is an important issue for this supernova. The SN~2003cg
spectra exhibit strong interstellar \NaI D lines at the rest
wavelength of the host, plus weaker absorption at the Milky Way
rest wavelength (Figure \ref{fig_naid}).  In addition, there is a
narrow absorption at $\sim$ 6282 $\AA$ not seen in SN~Ia spectra
(see Figure \ref{fig_naid}). This feature coincides with a
\textit{diffuse interstellar band} (DIB) measured in e.g.  HD
183143 at 6283.86 $\AA$ by \cite{herbig95}. It tends to be seen in
the spectra of stars which are heavily obscured and reddened by
interstellar dust, and is probably due to a variety of
interstellar polyatomic molecules based on carbon.

Adopting a Virgo distance of 15.3~Mpc \citep{freedman01} and a
relative distance from Virgo of 1.18 for NGC 3169 \citep{kraan86},
the host galaxy, we estimate a distance modulus for SN~2003cg of
$\mu = 31.28$.

Given the observed magnitude at maximum (see Section
\ref{light_curves}), we obtain an absolute V magnitude
$M_V^{max}\sim-16.56$ before any reddening correction. This
compares with $M_V^{max}=-19.46\pm0.05$ for a sample of nearby
dereddened SNe~Ia sample \citep{gibson00}. In addition, the
$(B-V)_{max}$ colour is $\sim$1.08 redder than typical SNe~Ia,
which have $(B-V)_{Bmax}\simeq0.00\pm0.04$ \citep{schaefer95},
again pointing to heavy reddening (see Section
\ref{colour_curves}).

From these various indicators we conclude that the light emitted
by SN 2003cg is heavily reddened. Thus, before discussing the
physical properties of the supernova, it is of utmost importance
to obtain a precise estimate of the reddening.

\subsection{Reddening estimation - Standard procedure} \label{reddstd}
In general, extinction is a function of wavelength. For the V~band
it is related to the colour excess \ebv~through the relation,

\begin{equation}
A_V=R_V \times E(B-V), \label{equ_redd}
\end{equation}

where \rv~is the ratio of total-to-selective absorption.
Therefore, the problem of estimating the extinction becomes that
of measuring the colour excess. Lira (1995) found that the
+30--+90~day SN~Ia \bv~colour curve is identical for all SNe,
independent of their light curve and in particular of their \dm15
(the B magnitude decline between the maximum and 15 days later,
\citealt{phillips93}). This indicates an \ebv~of 1.35 for
SN~2003cg. Using the standard value of \rv~= 3.1, we obtain an
intrinsic absolute magnitude of $M_V^{max}$ $\sim$ -20.91 cf. the
mean $M_V^{max}= -19.46\pm0.05$ obtained by \cite{gibson00}. This
implies that either this SN was exceptionally luminous, or that
the adopted \rv~is too large.\\

\rv~has been well determined in our Galaxy by comparing the colour
of reddened and unreddened stars of identical spectral type. Apart
from a small dispersion at the shorter wavelengths
(\citealt{seaton79}; \citealt{savage79}; \citealt{cardelli89}
(hereafter CCM); \citealt{odonnell94}; \citealt{calzetti01}),
there seems to be a remarkable homogeneity in the optical portion
of the interstellar extinction curve along various Galactic
directions, with a mean value of \rv~= 3.1.  However, for a few
directions values ranging from \rv~= $\sim$2 to $\sim$5.5 have
been found \citep{fitzpatrick04,geminale05}. Moreover, we have
relatively poor information about the extinction parameters in
other galaxies \citep{chini90,jansen94,clayton04}. In fact,
several SNe~Ia studies have yielded statistical evidence for lower
values of \rv: 0.70 $\pm$ 0.33 (\citealt{capaccioli90}), 0.5
(\citealt{branch92}), 2.3 $\pm$ 0.2 (\citealt{dellavalle92}), 2.5
$\pm$ 0.4 (\citealt{phillips99}), 1.8 (\citealt{krisc00}), 2.5
\citep{knop03,altavilla04}, 1.55 (\citealt{krisc05}), 1.1 and 3.1
(2 components) (\citealt{pozzo06}). Since \rv~is related to the
characteristics of dust clouds along the line-of-sight, SNe~Ia may
prove to be a useful probe of the grain properties.

\subsection{Reddening estimation - an anomalous extinction law} \label{methods_extinct}
In order to obtain a precise extinction estimate, we examine more
closely the relation between $A_{\lambda}$ and \ebv. CCM provided
analytic expressions for the average extinction law
$A_{\lambda}$/$A_V$ for the wavelength range $0.125 \mu m \leq
\lambda \leq 3.5 \mu m$. These are characterised by the parameter
\rv~= $A_V$ / \ebv. Their analytic formulae reproduce the Galactic
mean extinction curve \citep{seaton79,savage79} with \rv~close to 3.1. \\

\subsubsection{Photometric determination of the extinction} \label{methods_extinct_ph}
\begin{figure*}
\centering \psfig{figure=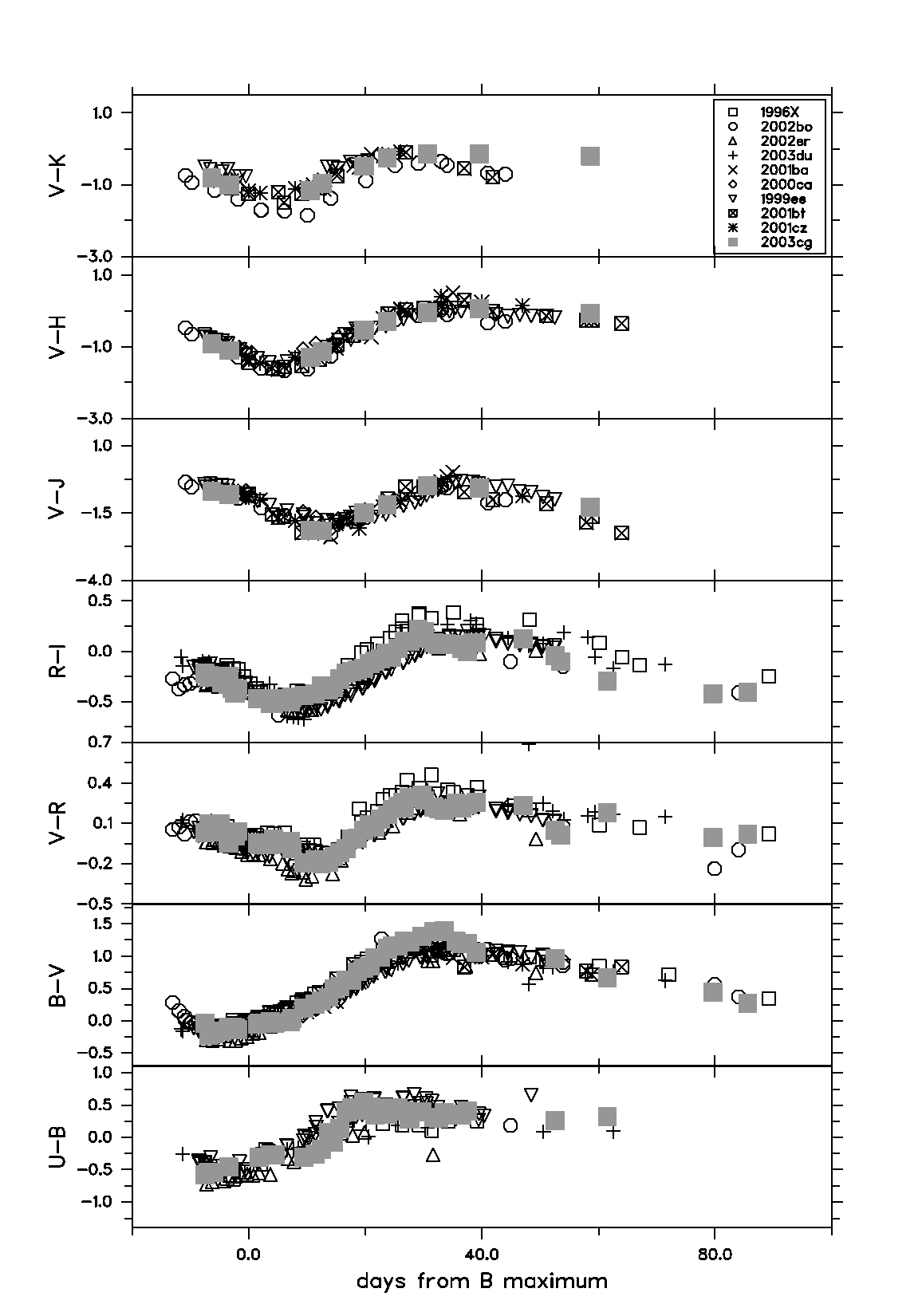,width=16.1cm,height=23cm}
\caption{Intrinsic optical and NIR colour curves of SN 2003cg
compared with those of SNe 2003du (Stanishev et al. in
preparation), 2002er (\citealt{pignata04}), 2002bo
(\citealt{benetti04}; \citealt{krisc04b}), 1996X
(\citealt{salvo01}), 2001bt, 2001cz (\citealt{krisc04b}), 1999ee,
2000ca and 2001ba (\citealt{krisc04a}). For a discussion on the
reddening adopted, see Section \ref{methods_extinct_ph}.}
\label{fig_color}
\end{figure*}

As indicated above, we have estimated the reddening of SN~2003cg
using Phillips et al.'s (1999) prescription based on the Lira
(1995) result for the late-time \bv~colour. We obtain \ebv~= 1.35
(see Table \ref{tabla_reddvalue}). However, in determining the
SN~2003cg reddening we would like to make use of most of the
colours available, not just (B-V). Therefore, we have compared a
range of optical and NIR colour curves of SN~2003cg with those of
other normal SNe~Ia (Figure \ref{fig_color}) (see also Section
\ref{colour_curves}). Assuming the CCM absorption law:

\begin{equation}
  \frac{A_{\lambda}}{A_{V}} = a_{\lambda} + \frac{b_{\lambda}}{R_{V}} ,
\label{equ_ccm}
\end{equation}

where $a_{\lambda}$ and $b_{\lambda}$ are wavelength-dependent
coefficients, we can write the following expression:


\begin{eqnarray}
  E({\lambda}_i-{\lambda}_j)& = & A_V \left [(a_{{\lambda}_i}-a_{{\lambda}_j}) +  \frac{b_{{\lambda}_i}-b_{{\lambda}_j}}{R_V}
  \right].
 \label{equ_colour}
\end{eqnarray}

The parameters of this equation were adjusted simultaneously to
provide the values of \ebv~and \rv~which gave the best (minimum
residual) simultaneous matches to the normal SN~Ia colour curves.
(cf. Figure \ref{fig_color}). This was achieved with \ebv~= 1.33
$\pm$ 0.28 and \rv~= 1.97 $\pm$ 0.29, clearly a smaller value than
the canonical \rv~= 3.1.

\begin{figure}
\begin{center}
\psfig{figure=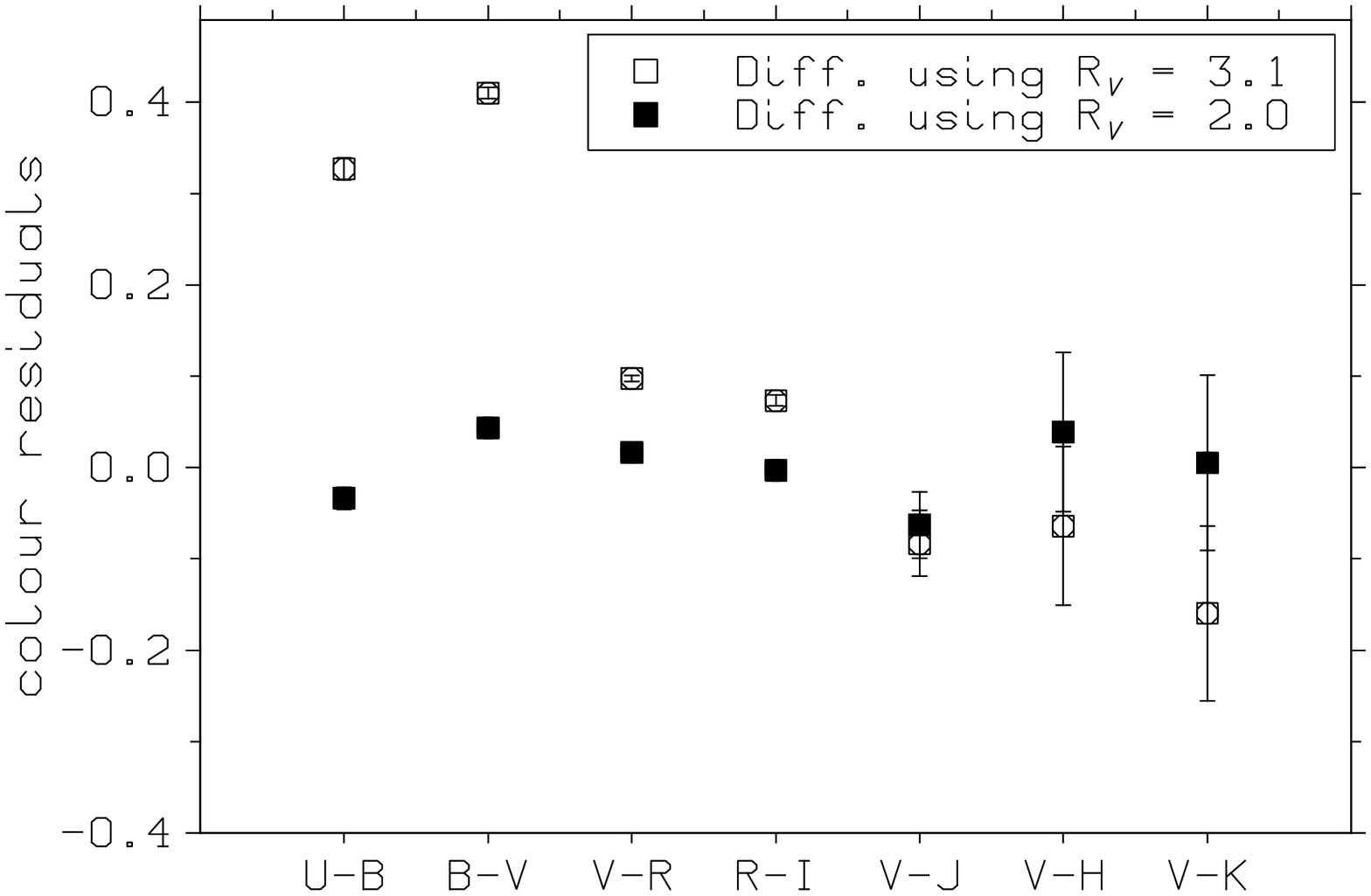,angle=270,width=9cm,height=6.4cm}
\caption{Differences between the observed colour excesses and
those computed for \rv = 3.1 (empty symbols) and \rv = 2.0 (full
symbols).} \label{fig_residuocolor}
\end{center}
\end{figure}

Figure \ref{fig_residuocolor} shows the average differences
between the observed colours and those expected for \rv~= 3.1 and
\rv~= 2.0 ($\equiv$ 1.97). This plot confirms our result that
\rv~is small: indeed for \rv~= 2.0 the residuals are practically
zero.

\subsubsection{Spectroscopic determination of the extinction} \label{methods_extinct_spec}
An independent method for determining the extinction is via the
equivalent width (EW) of the narrow interstellar \NaI D doublet.
This is believed to be related to the amount of dust between us
and the source (\citealt{barbon90} and \citealt{munari97}).
\cite{turatto03} studying a sample of SNe~Ia, found that the
points appear to cluster around two separate relations between EW
(\NaI D) and \ebv:

\begin{eqnarray}
\label{equ_naid1}
E(B-V) = 0.16 \times EW (NaI D)\\
\label{equ_naid2}
E(B-V) = 0.51 \times EW (NaI D)\\
\nonumber
\end{eqnarray}

Averaged over the 25 spectra obtained for SN~2003cg, the \NaI D
doublet has an EW = 0.63 $\pm$ 0.07 $\AA$ for the Milky Way and EW =
5.27 $\pm$ 0.50 $\AA$ for NGC 3169. Using the relations
(\ref{equ_naid1}) and (\ref{equ_naid2}) we find \ebv $\sim$ 0.10 and
\ebv $\sim$ 0.32 for the Milky Way, and \ebv $\sim$ 0.84 and \ebv
$\sim$ 2.69 for NGC~3169. The sum of the lower values (from relation
\ref{equ_naid1}) for the host galaxy and the Milky Way is 0.94 which
is about 0.4 lower than those found in Section
\ref{methods_extinct_ph}. However, since the ratio between the \NaI D
component lines is close to saturation (1.24), we regard this value of
\ebv as being a lower limit only.  We also considered using the EW of
the K\,{\sc i} (7699$\AA$) absorption line to estimate the reddening
\citep{munari97}. Unfortunately, the spectra covering this spectral
region were of insufficient resolution to detect this much weaker
line.

Another way of determining \ebv~and \rv~is to use the optical SED
(3400-9350$\AA$) of SN~2003cg. We have developed a script which
derives the free parameters of a CCM extinction law by comparison
of the SED of SN~2003cg with those of unreddened SNe~Ia, at
various coeval epochs. The reference SNe were selected to have
similar light curve shapes (\dm15) and spectral features. Their
spectra were corrected for redshift and Galactic reddening. The
reference spectra were then scaled to the distance of SN 2003cg
via:

\begin{equation}
  f_{ref}^{03cg} = (\frac{d_{ref}}{d_{03cg}})^2 f_{ref},
\label{equ_scrip1}
\end{equation}

where $f_{ref}$ is the observed flux of the reference SN at its
true distance, $d_{ref}$, $d_{03cg}$ are the distances of the
comparison SN and SN 2003cg respectively and $f_{ref}^{03cg}$ is
the flux of the reference SN at the distance of SN 2003cg. The
script divides the SN~2003cg spectra by the scaled reference
spectra and calculates:

\begin{equation}
  A_{\lambda} = -2.5 log \frac{f_{03cg}}{f_{ref}^{03cg}},
\label{equ_scrip2}
\end{equation}

where $A_{\lambda}$ is the total extinction at any wavelength and
$f_{03cg}$ is the observed flux of SN 2003cg.  Finally, in order
to compare with the CCM law (relation \ref{equ_ccm}), we normalise
the derived extinction curve to match the value of $A_{\lambda}$
at the V band effective wavelength.

\begin{figure*}
\begin{center}
\psfig{figure=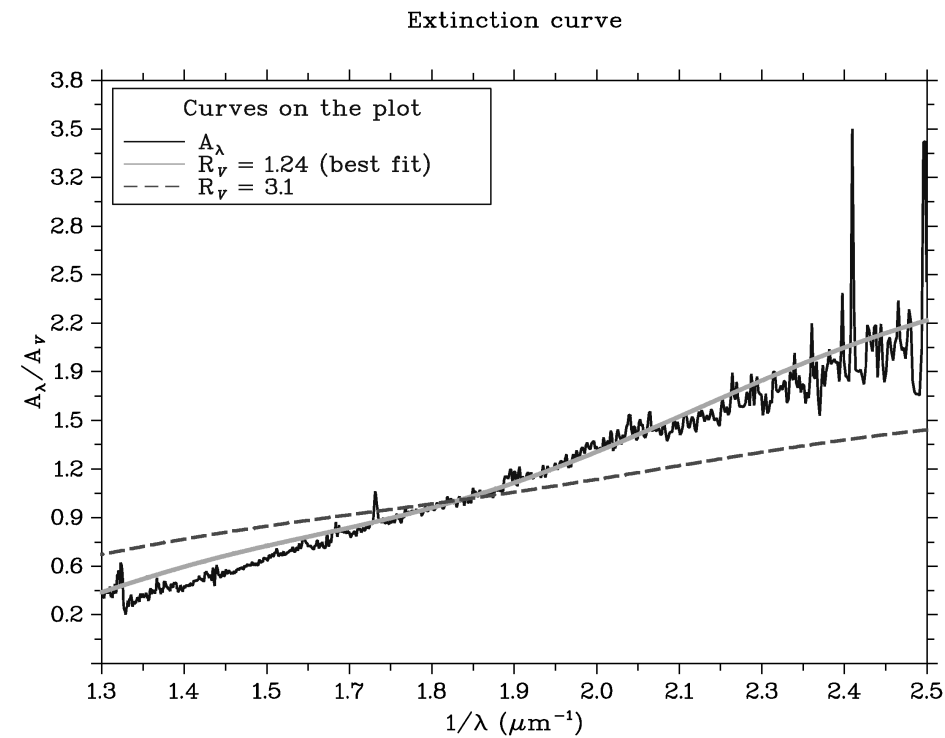,angle=270,width=16.5cm,height=11.67cm}
\psfig{figure=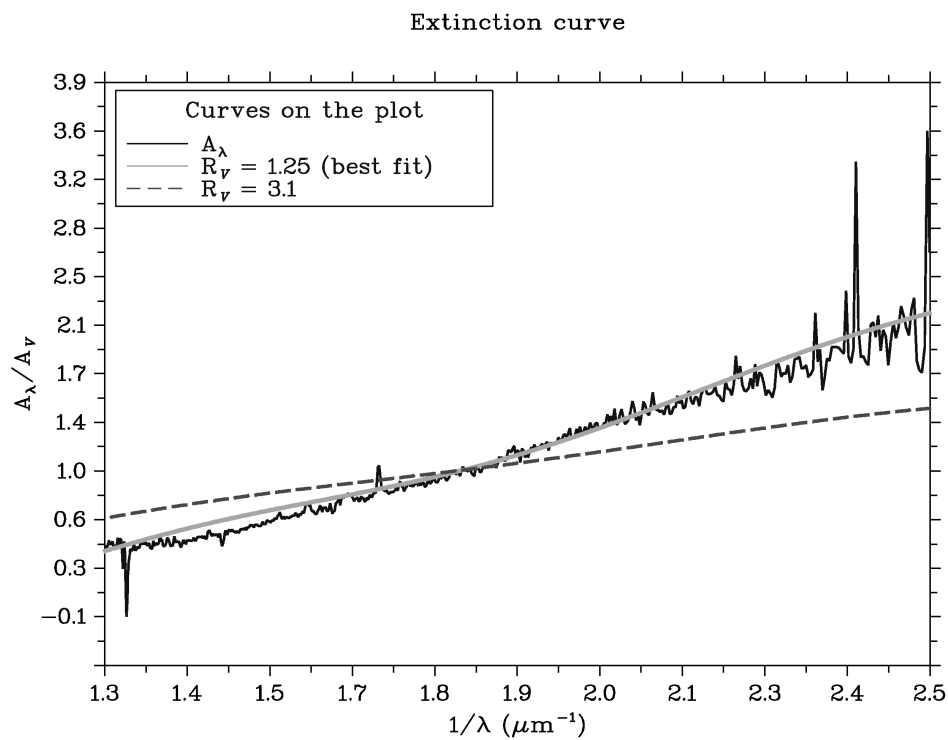,angle=270,width=16.5cm,height=11.67cm}
\caption{Best fit of theoretical CCM laws (solid line) to
empirical extinction curves of SN 2003cg obtained with the pair
1994D-2003cg (top) and 1996X-2003cg (bottom) on day -2. The CCM
extinction curve for \rv = 3.1 (dashed line) is plotted for
comparison.} \label{fig_card}
\end{center}
\end{figure*}

As reference we have used the spectra of SN~1994D (\dm15 = 1.32 -
\citealt{patat96}, \citealt{phillips99}) and SN 1996X (\dm15 =
1.31 - \citealt{salvo01}). We have compared seven pairs of
94D-03cg spectra at different epochs and five pairs of 96X-03cg
spectra, obtaining the parameters $A_{V}$ and \rv~in each case.
The average values of these parameters for each sets of pairs are
given in Table \ref{tabla_reddvalue}. Two examples of this
comparison are
shown in Figure \ref{fig_card}.\\

From the \ebv\, and \rv\, values listed in Table
\ref{tabla_reddvalue} (but excluding the \NaI D measurement) we
obtained weighted average values of \ebv = 1.33 $\pm$ 0.11 and
\rv~= 1.80 $\pm$ 0.20. Using these results we can estimate the NIR
colour excesses using the basic relations between $A_{V}$,
\ebv~and \rv: $E(V-J)$ = 1.92, $E(V-H)$ = 2.08 and E($V-K$) =
2.20. These values are approximately 10\% smaller than those we
obtained from the comparison of the colour curves reported on
Table \ref{tabla_mainphdata} (see also Section \ref{methods_extinct_ph}).\\

Up to this point we have investigated the extinction using methods
which either did not involve the value of \dm15 or where the
reference SNe~Ia were of similar light curve shape. If a wider
diversity of SNe~Ia were employed we would have to correct this
parameter for the effects of reddening (see Section
\ref{earl_ph}). Assuming $\Delta m_{15}(B)_{intrinsic}$ = 1.25, we
checked the \ebv~using the prescription given by \cite{reindl05}
at maximum for a sample of 111 "Branch normal" SNe Ia and
tentatively for five 1991T-like and +35 days for 59 normal SNe Ia.
We obtained for the two epochs \ebv~= 1.21 $\pm$ 0.20 and 1.25
$\pm$ 0.11 respectively, which are in excellent
agreement with the estimates reported before.\\

This analysis confirms that the light emitted by SN 2003cg was heavily
reddened. However, in order to match its intrinsic luminosity and
colour evolution to those of other SNeIa with similar light curve
shapes and spectral features, it was necessary to deredden the
SN~2003cg data using a value of \rv=1.80 $\pm$ 0.20. This is
significantly less than the standard ISM value of 3.1.

\subsection{Effects of a light echo?} \label{possib_echo}
Wang (2006) has suggested that the anomalous reddening seen
towards a number of SNe~Ia may be due to the effects of a dusty
circumstellar cloud. In this ``light-echo'' (LE) scenario, not
only is light scattered out of the beam, but also we see light
that has been scattered into our line-of-sight. Thus, in addition
to the scattering cross-section we also need to consider the
(wavelength-dependent) albedo of the grains. Wang finds that a
dust cloud of inner radius $10^{16}$cm would produce a significant
reduction in \rv. However, we suggest that there are a number of
difficulties with this scenario. Firstly, the survival of grains
at such a small distance from the supernova is rather unlikely.
Adopting the bolometric luminosity of SN~2003cg derived below, and
scaling from Dwek's (1983) analysis of SN~1979C we find that the
inner radius of a dust cloud around SN~2003cg would exceed
$\sim2\times10^{17}$cm. Dust at this large distance would have
only a modest effect on \rv~(see Wang (2--6) Figure 4). In
addition, a nearby, dusty CSM would be expected to produce other
observable effects at later times.  By 50--100~days, the
fastest-moving ejecta would collide with the CSM producing
characteristic spectral features as seen in the peculiar Type~Ia
SN~2002ic \citep{hamuy03}. However, no signs of such an
interaction are apparent.  Moreover, even if some (larger) grains
survived the peak luminosity, their high temperature would produce
a strong near-IR excess, as was seen in SN SN~2002ic
\citep{kotak04}. However, no normal SN~Ia, including SN~2003cg,
has shown such an effect.  Other LE effects expected but not seen
in SN~2003cg, include (a) a reduced $(B-V)$ color range (so, that
after matching the colors at maximum, at later phases one would
expect the observed color to be bluer than that of a LE-free
object), (b) temporal variation in the reddening, (c) an
anomalously small $\Delta m_{15}$, (d) a significantly brighter
late phase tail and (e) broader spectral lines \citep{patat05}. We
conclude that any circumstellar dust, if present, has only a small
effect on
the extinction behaviour of SN~2003cg.\\

We conclude that the small value of \rv~for SN~2003cg is most likely
due to a grain size distribution where the grains are generally
smaller than in the local ISM \citep{geminale06}.  A more detailed
analysis of possible LE effects will be presented in a separate
paper.\\

In the photometric and spectroscopic analysis of SN~2003cg that
follows, we adopt \ebv = 1.33 $\pm$ 0.11 and \rv = 1.80 $\pm$
0.20.

\begin{table*}
\centering \caption{Values of the colour excess and the ratio of
total-to-selective extinction derived from different
methods.}\label{tabla_reddvalue}
\begin{tabular}{cccl}
\hline
 Method           &   \ebv      &  \rv &  Reference\\
\hline
$\ebv_{tail}$     & 1.35$\pm$0.13 &      --     & \cite{phillips99} \\
Colour Evolution  & 1.33$\pm$0.28 &1.97$\pm$0.29& Section \ref{methods_extinct_ph}\\
Comp-CCM (94D)    & 1.31$\pm$0.44 &1.61$\pm$0.41& Section \ref{methods_extinct_spec} \\
Comp-CCM (96X)    & 1.25$\pm$0.33 &1.69$\pm$0.33& Section \ref{methods_extinct_spec}  \\
EW of \NaI D      & 0.94          &      --     & \cite{turatto03}\\
 \hline
\end{tabular}
\end{table*}

\section{Photometry} \label{ph_results}

\subsection{Light curves} \label{light_curves}
The light curve shape provides one of the main ways by which we
acquire information about individual SNe. It is linked directly to
the energy input and to the structure of the exploding star.
However, it can also be influenced by intervening dust. In the
following sections we discuss the early and late light curve
phases of SN~2003cg.

\subsubsection{Early phase} \label{earl_ph}
\begin{figure*}
\begin{center}
\psfig{figure=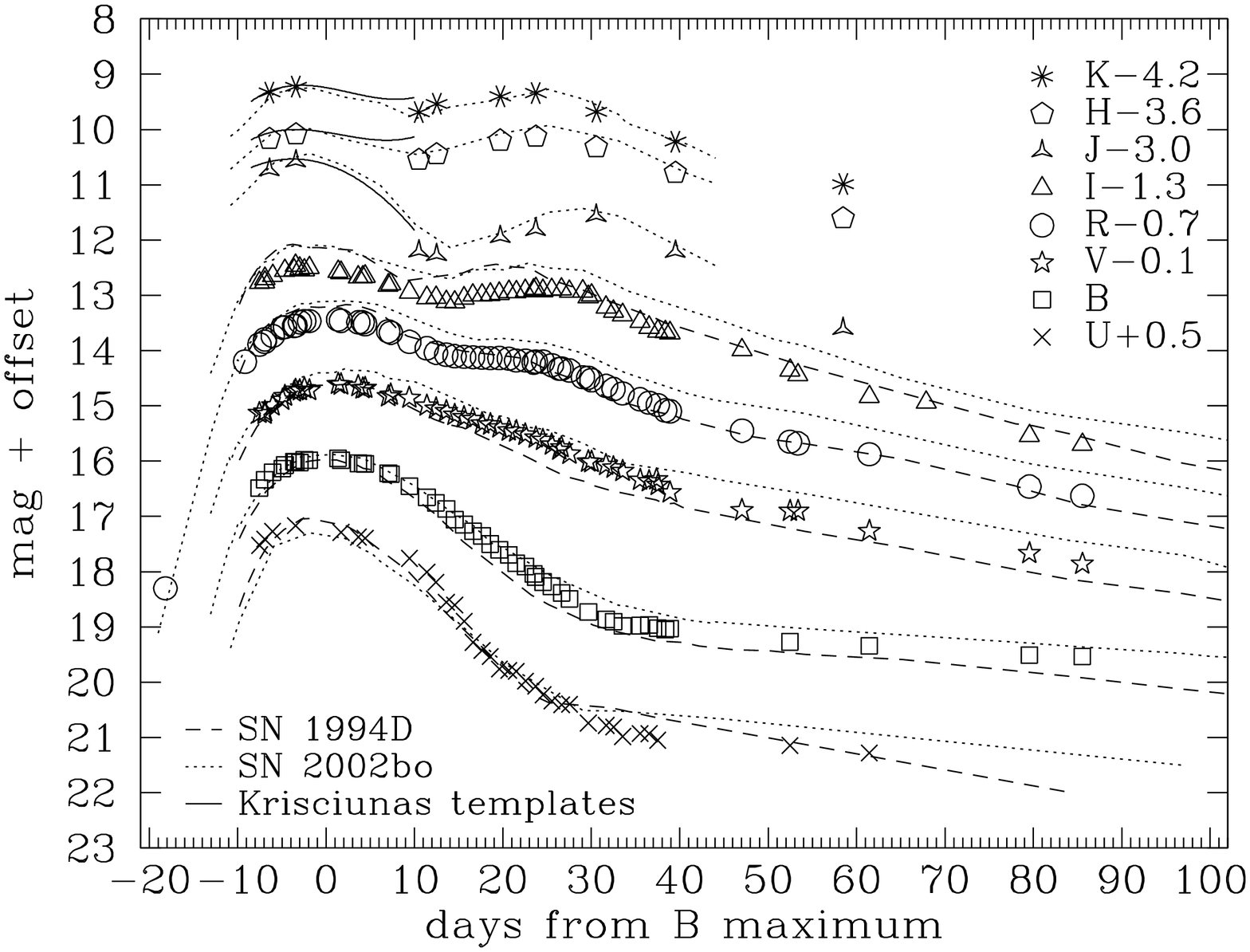,angle=270,width=\textwidth}
\caption{S-corrected optical and IR light curves of SN 2003cg
during the first months post-explosion. The light curves have been
shifted by the amount shown in the legend. The dashed, dotted and
solid lines represent the light curves of SN 1994D
(\citealt{patat96}), SN 2002bo (\citealt{benetti04},
\citealt{krisc04b}) and JHK templates (\citealt{krisc04a}), respectively, adjusted to the SN 2003cg
distance and reddening (\ebv = 1.33 $\pm$ 0.11 and \rv = 1.80
$\pm$ 0.20).
}
\label{fig_lightcurv}
\end{center}
\end{figure*}

The optical photometry (Table \ref{tabla_ori_ph_opt},
\ref{tabla_scorr}) \& and near-IR photometry (Table
\ref{tabla_ori_ph_ir}) of SN 2003cg are plotted as light curves in
Figure \ref{fig_lightcurv}.  Note in particular the excellent sampling
of the optical data from --7 to +85 days.  Also shown as ``$R$-band''
data are the unfiltered pre-discovery and discovery points of
\cite{itagati03} (see below). In view of the small redshift of
SN~2003cg (z=0.004), we have not applied K-corrections to any of the
magnitudes. The shapes of the light curves are typical of a normal
Type~Ia SN e.g. (a) the occurrence of secondary maxima longward of the
V-band (b) the increased prominence of the secondary maxima in the
NIR.  Also typical is that maximum light occurred before $t_{Bmax}$ in
the U, I and NIR bands, and slightly after in the V and R bands
(e.g. \citealt{contardo00}).  In order to estimate the maximum light
epochs and magnitudes for the different bands, we fitted low order
polynomials to the light curves around their respective maxima. The
results are given in Table \ref{tabla_mainphdata} together with other
parameters for SN 2003cg and its host galaxy.\\

An important light curve width/shape parameter is \dm15. For
SN~2003cg we obtain a direct measurement of \dm15 = 1.12 $\pm$
0.04. However, the intrinsic light curve shape may be altered by
the combined effects of SED evolution and dust extinction. We
corrected for this effect using the formula of \cite{phillips99}
viz. correction = 0.1$\times$\ebv. For the \ebv~derived in the
previous section this gives a correction of 0.133 and hence an
intrinsic value of \dm15 = 1.25 $\pm$ 0.05. However, in view of
the high reddening of SN 2003cg, we decided to check the \dm15
correction using synthetic magnitudes obtained from its spectra.
We computed \dm15 for (a) the observed spectra and (b) the spectra
corrected for Galactic and host galaxy reddening (see Sections
\ref{indredd} and \ref{spec_results}). The difference between the
two synthetic values is 0.128, very similar to that obtained from
the Phillips et al. formula, thus indicating that their formula is
valid even at high reddening.

Another useful means of characterising the light curve shape is
the stretch factor s \citep{perlmutter97,goldhaber01,altavilla04}.
This is the factor by which an observed $B$-band light curve must
be expanded or contracted to match a standard light curve shape.
We find s = 0.97 $\pm$ 0.02 for SN 2003cg, in good agreement with
that obtained via the relation \dm15 = $1.98\times(s^{-1}-1)+1.13$
\citep{altavilla04}.  \\

In Figure \ref{fig_lightcurv}, for comparison we also plot the
light curves of two other nearby Type Ia SNe having similar \dm15:
SN 1994D (\dm15 = 1.32, \citealt{phillips99}) and SN 2002bo (\dm15
= 1.13, \citealt{benetti04}). Both SNe have been: (a) shifted to
the SN 2003cg distance (Section \ref{bolometric}), (b) dereddened
according to their colour excesses (0.04 - \citealt{phillips99}
and 0.38 - \citealt{stehle05} respectively) with \rv = 3.1, and
(c) artificially reddened with the \ebv~and \rv~values of SN
2003cg (Section \ref{indredd}). We adopt distance modules of 30.68
and 31.67, for SN 1994D and SN 2002bo, respectively
\citep{patat96,benetti04}.

With respect to the other two SNe, SN~2003cg has a broader U-band
peak.  This can be due to the high reddening towards SN~2003cg
which shifted the bandpass to redder effective wavelengths where
the light curves are broader. The same effect is probably
responsible for the pronounced shoulder in the V band light curve
which correlates with the emergence of the secondary maxima in the
light curves in the R, I and longer wavelength light curves.

SN 1994D and SN 2003cg show similar maximum magnitudes in the UBV
bands but in R and I SN 1994D is brighter. In addition, the secondary
maximum in I occurs approximately 8 days earlier in SN 1994D. SN
2002bo is approximately 0.5 magnitudes brighter than SN 2003cg in the
VRI bands, and approximately 0.5 magnitudes fainter in the U band. In
contrast with the optical bands, we find similar behaviour between SN
2003cg and SN 2002bo in the NIR bands (no NIR light curves are
available for SN~1994D).  We note that the H and K light curve minima
of both SN~2002bo and 2003cg occur later and are more pronounced than
in the Krisciunas et al. NIR template light curves
(\citealt{krisc04a}).  \\

As already mentioned (Section \ref{obs}), SN 2003cg was discovered
about 10 days before B maximum. Along with the discovery,
\cite{itagati03} report two unfiltered optical images
obtained at --18.1 and --9.1 days (see Table
\ref{tabla_ori_ph_opt}). We judge that the magnitudes from these
images most closely approximate to those which would have been
obtained in the R~band \citep{pignata04}.

\cite{riess99} suggest that the early ($t_0 \la 10$ days)
luminosity is proportional to the square of the time since
explosion and derived an estimate of the explosion epoch $t_0$.
Applying this relationship to the SN~2003cg R~band light curve
(including the two unfiltered points), we obtain $t_0$(R) $\la$
52709.5 $\pm$ 0.6 (MJD) i.e. a risetime of 20.4 $\pm$ 0.5 days.
However, for the B band is $t_{r}$(B) = 19.9 $\pm$ 0.5 days,
significantly longer then that derived by \citep{riess99} (17.71
days) for a SN Ia with a \dm15 similar to that of SN 2003cg.

\subsubsection{Nebular phase} \label{neb_ph}
\begin{figure*}
\begin{center}
\psfig{figure=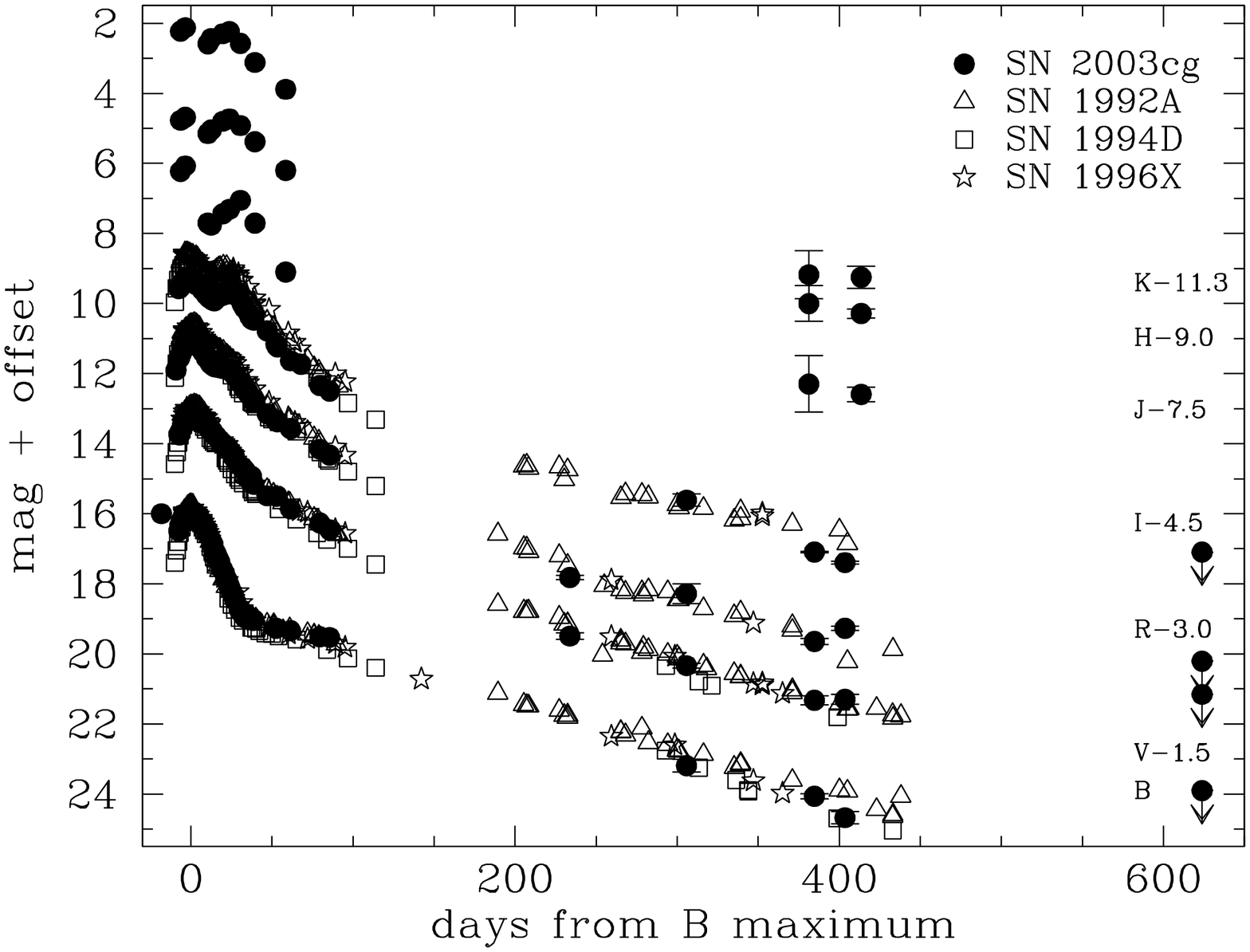,angle=270,width=\textwidth}
\caption{S-corrected BVRIJHK light curves for SN 2003cg covering
early and late phases. For clarity, the light curves have been
displaced vertically by the amounts shown on the right. Triangles,
squares and stars represent the data of SN 1992A
\citep{suntzeff96,altavilla04}, SN 1994D \citep{patat96,
altavilla04} and SN 1996X (\citealt{salvo01}) respectively. They
have been adjusted to the SN 2003cg distance and reddening (\ebv =
1.33 $\pm$ 0.11 with \rv = 1.80 $\pm$ 0.20). The lack of data from
3 to 7 months, and from 13 to 20 months is due to seasonal gaps.}
\label{fig_lightcurvneb}
\end{center}
\end{figure*}

The optical (BVRI) and NIR evolution of SN~2003cg was also
monitored at late phases (to over one year post-maximum). The
complete light curves, covering both early and late phases, are
shown in Figure \ref{fig_lightcurvneb}, together with those of SNe
1992A \citep{suntzeff96,altavilla04}, 1994D
\citep{patat96,altavilla04} and 1996X (\citealt{salvo01}). The
evolution of the SN~2003cg optical fluxes is consistent with
exponential declines in each band. The decline rates are given in
Table \ref{tabla_latedecline}. The late time optical evolution of
the four SNe~Ia is broadly similar. The BVRI light curve declines
are consistent within the errors with those estimated by
\cite{turatto90} for SNe~Ia. In the NIR, only two points quite
close in time are available, and the errors are large.
Consequently, little can be said about the NIR decline rate (Table
\ref{tabla_latedecline}).

\begin{table*}
\centering \setlength\tabcolsep{3pt} \caption{Decline rates of SN
2003cg at late phases$^1$.}\label{tabla_latedecline}
\begin{tabular}{ccccccc}
\hline
B            &   V      &  R & I & J & H & K \\
\hline 1.40$\pm$0.31 & 1.24$\pm$0.14 & 1.23$\pm$0.10 &
1.99$\pm$0.24 & 0.91$\pm$0.51 & 0.89$\pm$0.32 & 0.22$\pm$0.51 \\
\hline
\end{tabular}
\begin{flushleft}
$^1$ Magnitude per 100 days between 233 and 385 days for optical data, and between 381 and 414 for NIR data.\\
\end{flushleft}
\end{table*}

\subsection{Colour curves} \label{colour_curves}

We have already demonstrated that SN~2003cg is a heavily extinguished
SN~Ia. In Section \ref{indredd} we discussed extensively the
reddening of SN 2003cg and the use of the colour as an extinction
indicator. In this section we discuss only the intrinsic colour
evolution of the supernova.

\begin{figure}
\begin{center}
\psfig{figure=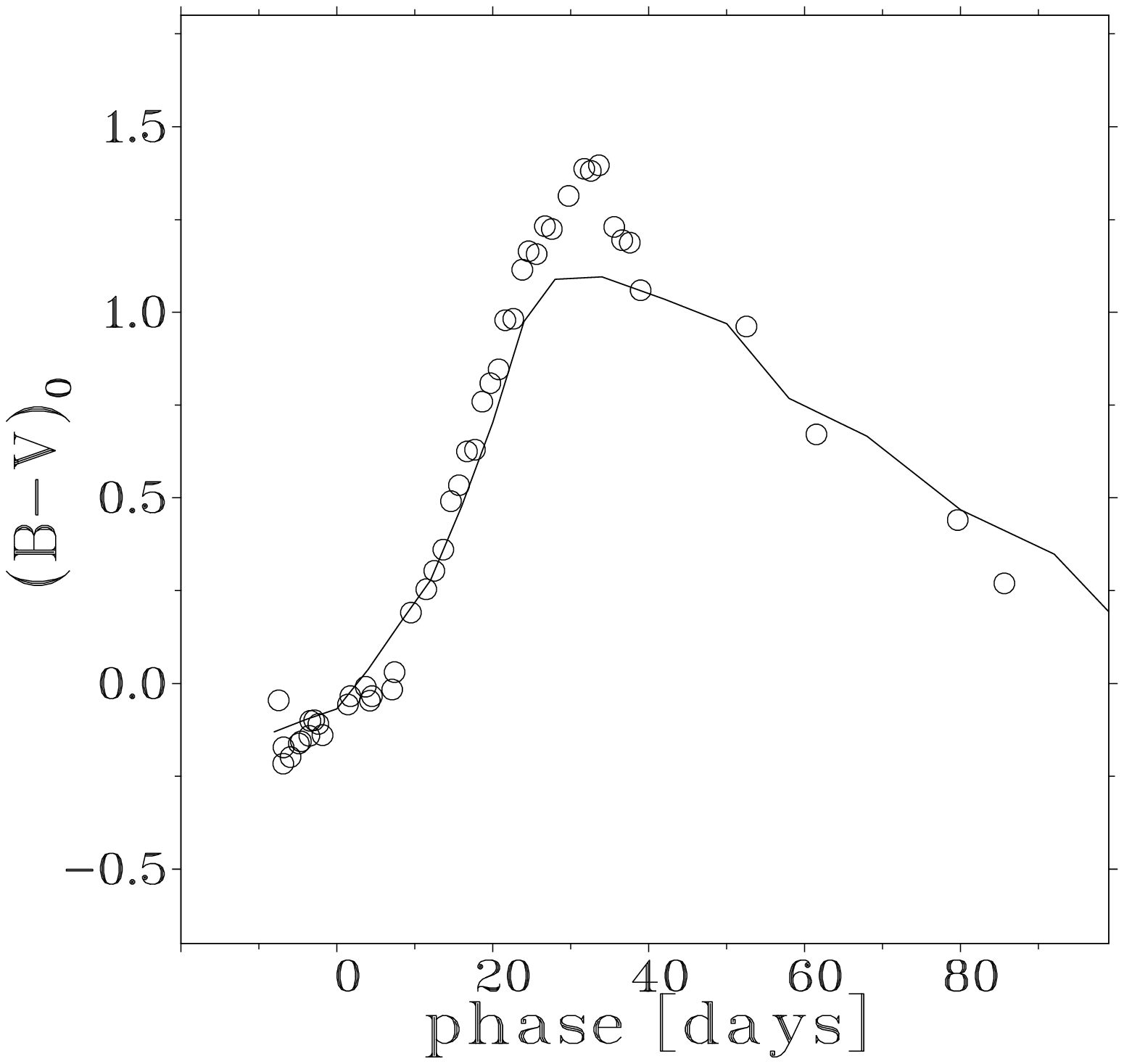,width=8cm,height=8cm}
\caption{Intrinsic (B-V) colour curve of SN 2003cg (circles)
compared with the average $(B-V)_0$ colour curve (solid line)
obtained from SNe 2003du (Stanishev et al. in preparation), 2002er
\citep{pignata04}, 2002bo \citep{benetti04,krisc04b}, 1996X
\citep{salvo01}, 2001bt, 2001cz \citep{krisc04b}, 1999ee, 2000ca
and 2001ba \citep{krisc04a}. For a discussion on the reddening
adopted, see Section
\ref{methods_extinct_ph}.}\label{fig_colorsolobv}
\end{center}
\end{figure}

In Figure \ref{fig_color}, the intrinsic colour curves of
SN~2003cg (corrected for \ebv = 1.33 $\pm$ 0.11 and \rv = 1.80
$\pm$ 0.20) are compared with those of other SNe Ia (cf. Section
\ref{indredd}). The optical colour curves are generally very
similar. The most significant difference is in the $(B-V)_0$
colour around $+30^d$ when SN~2003cg presents a redder bump
reaching a $(B-V) \sim 1.4$ (see also Figure
\ref{fig_colorsolobv}). At maximum, SN~2003cg has $(B-V)_0 \sim
-0.28$, and curiously, \cite{phillips99} and \cite{jha02} found
$(B-V)_{Bmax}$ values of --0.09 and --0.10, respectively, for
normal SNe Ia (after correction for Galactic and host galaxy
extinctions). Consequently, SN 2003cg would stand out from the
majority of points on the colour-colour diagram ($(U-B)_{max}$ vs.
$(B-V)_{max}$) of SNe Ia shown in figure 3.9 of \cite{jha02}, and
so, in our case, his discussion is not applicable.

The $V-J$, $V-H$ and $V-K$ colour curves of SN 2003cg are also shown
in Figure \ref{fig_color} (top panels). Again, the behaviour of
SN~2003cg is very similar to that of the comparison sample. In
particular, the IR colour evolution of SN 2003cg matches well that of
SN 2001bt (\citealt{krisc04b}). \\

In a recent work \cite{wang05} introduced the colour parameter,
$\Delta C_{12}$, the (B-V) colour measured 12 days after the B
maximum. This parameter depends strongly on the decline rate \dm15
and may provide a means of estimating the reddening due to the
host galaxy. They also find a correlation between the peak
luminosities of SNe Ia and $\Delta C_{12}$. For
SN 2003cg, we measure $\Delta C_{12}$ = 0.39 $\pm$ 0.11.\\

In conclusion we emphasise that, given the reddening derived in
Section \ref{indredd}, the intrinsic colour evolution of SN 2003cg
is normal over a wide range of wavelengths.

\subsection{Absolute magnitudes and the {\it uvoir} light curve} \label{bolometric}
For a Virgo distance of 15.3 Mpc \citep{freedman01} and a relative
distance from Virgo of 1.18 for NGC 3169 \citep{kraan86}, the host
galaxy of SN 2003cg, we estimate a distance modulus of $\mu =
31.28$. Using the values for \ebv~and \rv~discussed in Section
\ref{indredd}, we obtain $M_B^{max}$ = -19.17 $\pm$ 0.75 and
$M_V^{max}$ = -19.00 $\pm$ 0.49 which is typical for SNe~Ia of its
\dm15. We have also used distance-independent parameters to
estimate $M_B^{max}$. In Table \ref{tabla_absolmag} we show the
SN~2003cg $M_B^{max}$ values derived using peak luminosity v.
\dm15 relations \citep{hamuy96,phillips99,altavilla04,reindl05}
scaled to $H_0$ = 72 \kms $Mpc^{-1}$ \citep{freedman01}. We also
give $M_B^{max}$ obtained from the parameter $\Delta C_{12}$
\citep{wang05}. The weighted average $M_B^{max}$ = -19.00 $\pm$
0.06 is consistent with the $M_B^{max}$ value obtained from the
observed peak magnitude and adopted distance modulus. ($M_B^{max}$
deduced from the \cite{reindl05} relation is somewhat discrepant).

A similar comparison using the stretch factor, s, can be made.
SN~2003cg has a stretch factor of 0.97 i.e. almost 1, and so we
can directly compare its absolute magnitude of $M_B^{max}$ =
-19.17 $\pm$ 0.75 with the $M_B^{max}$ = -19.30 value for s = 1
\citep{nugent02,knop03}. Again the agreement is good.

In contrast with the optical region, the IR magnitudes at maximum
($M_J^{max}$ = -18.22 $\pm$ 0.14, $M_H^{max}$ = -17.92 $\pm$ 0.12
and $M_K^{max}$ = -18.08 $\pm$ 0.05) are between 0.4 and 0.5 mag
fainter than the values found by \cite{meikle00} and
\cite{krisc04b}
for two samples of Type Ia SNe. \\

\begin{figure*}
\begin{center}
\psfig{figure=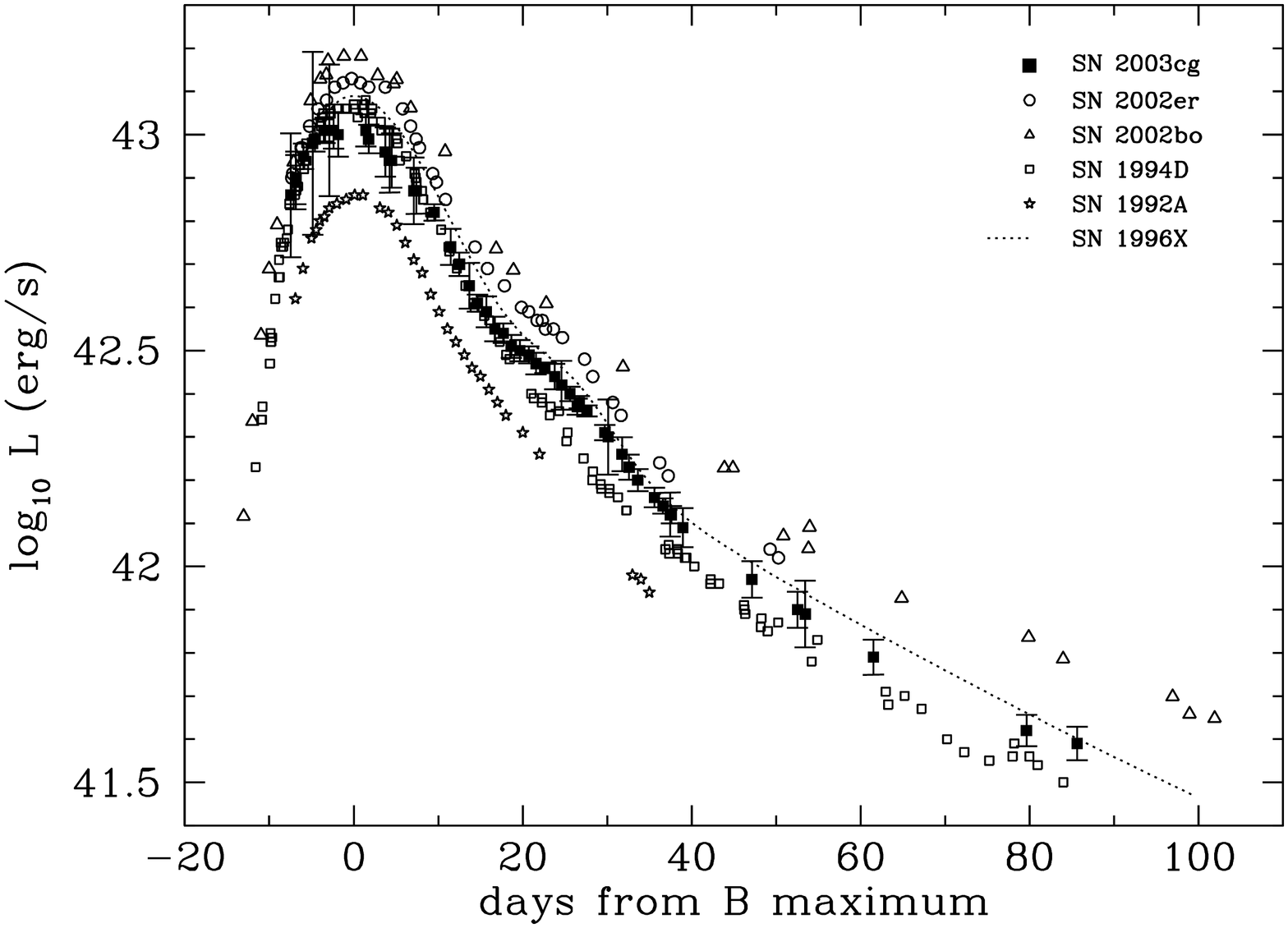,angle=270,width=\textwidth}
\caption{{\it uvoir} light curve for SN 2003cg. Filled squares
give the {\it uvoir} light curve of SN 2003cg. Dotted line is the
{\it uvoir} light curve for SN 1996X. Open stars, squares, circles
and triangles give the {\it uvoir} light curve for SN 1992A, SN
1994D, SN 2002er and SN 2002bo, respectively. Error bars refer
only to photometric errors and not to uncertainties in reddening
and distance.} \label{fig_bolometric}
\end{center}
\end{figure*}

The "bolometric" ({\it uvoir}) luminosity evolution of SN 2003cg is
shown in Figure \ref{fig_bolometric}. It was obtained using the
computed distance modulus and reddening, integrating the fluxes in the
UBVRIJHK bands (we used a combination of optical and IR spectra having
similar epochs) and adding the correction of \cite{suntzeff96} for the
unknown ultraviolet contribution. The reddening-corrected bolometric
luminosity at maximum is log$_{10}L$ = 43.01 $\pm$ 0.05 \ergs. Also
shown in Figure \ref{fig_bolometric} are the {\it uvoir} light curves
for five other SNe~Ia. The {\it uvoir} light curve shape of SN 2003cg
is clearly similar. In particular, SNe~1996X, 2002er and 2003cg all
exhibit prominent bumps around +25~days corresponding to the second
NIR maxima.  However, SN 2003cg is fainter than SN 1996X around
maximum. Differences in the bolometric luminosities at maximum may be
due to errors in the distance modulus or reddening estimates for the
different SNe.

\begin{table}
\centering \setlength\tabcolsep{3pt} \caption{Absolute B magnitude
of SN 2003cg derived from different
methods.}\label{tabla_absolmag}
\begin{tabular}{cll}
\hline
B            & Method &  Reference \\
\hline
-19.17$\pm$0.75  & Distance & This work \\
\hline
-18.91$\pm$0.13 & $M_B^{max}$ vs. \dm15$^1$ & \cite{hamuy96}$^2$    \\
-18.90$\pm$0.10 & $M_B^{max}$ vs. \dm15 & \cite{phillips99}   \\
-19.21$\pm$0.12 & $M_B^{max}$ vs. \dm15 & \cite{altavilla04}$^3$  \\
-18.22$\pm$0.56 & $M_B^{max}$ vs. \dm15 & \cite{reindl05}     \\
-19.20$\pm$0.22 & $M_B^{max}$ vs. $\Delta C_{12}$& \cite{wang05} \\
\hline
\end{tabular}
\begin{flushleft}
($1$) $\Delta m_{15}(B)_{intrinsic}$;\\
($2$) according to the relation given in Table 3 of \cite{hamuy96}
(peak luminosity);\\
($3$) according to the relation given in Table 1 of
\cite{altavilla04} ($\Delta$Y/$\Delta$Z=2.5).
\end{flushleft}
\end{table}

\begin{table}
\centering \setlength\tabcolsep{3pt}
\caption{Main data of SN
2003cg and its host galaxy.}\label{tabla_mainphdata}
\begin{tabular}{lll}
\hline
 Host Galaxy Data & NGC 3169 &  Ref.\\
\hline
$\alpha$ (2000)   & $10^{h}14^m 15\psec00$ & 1 \\
$\delta$ (2000)   & $03^{\circ} 27\arcmin 58\arcsec$& 1\\
Galaxy type       & SA(s)a pec & 1\\
B magnitude       & 11.08& 1 \\
Galactic extinction $A_B$ & 0.134 & 2\\
$v_{r}$ (\kms)    & 1238 & 1\\
$\mu$             & 31.28 & 3 \\
\hline
\hline
 SN Data          & SN 2003cg  &  Ref.\\
\hline
$\alpha$ (2000)   & $10^{h}14^m 15\psec97$ & 4 \\
$\delta$ (2000)   & $03^{\circ} 28\arcmin 02\parcsec50$& 4 \\
Offset SN-Gal. nucleus & $14\arcsec E, 5\arcsec N$ & 4 \\
Discovery date (UT)  & 2003 March 21.51 & 4 \\
Discovery date (MJD) & 52719.51 & 4 \\
\ebv              & 1.33 $\pm$ 0.11 & 5 \\
\rv               & 1.80 $\pm$ 0.19 & 5 \\
Date of B max (MJD)& 52729.40 $\pm$ 0.07 & 5 \\
Magnitude and epoch & U = 16.64 $\pm$ 0.03;  -1.6 (days) & 5 \\
at max respect B max & B = 15.94 $\pm$ 0.04;   0.0 (days) & 5 \\
                   & V = 14.72 $\pm$ 0.04;  +0.8 (days) & 5 \\
                   & R = 14.13 $\pm$ 0.05;  +0.5 (days) & 5 \\
                   & I = 13.82 $\pm$ 0.04;  -1.7 (days) & 5 \\
                   & J = 13.55 $\pm$ 0.05;  -2.6 (days) & 5 \\
                   & H = 13.69 $\pm$ 0.05;  -0.2 (days) & 5 \\
                   & K = 13.40 $\pm$ 0.02;  -0.7 (days) & 5 \\
Magnitude and epoch & I = 14.19 $\pm$ 0.02;  +26.4 (days) & 5 \\
of second IJHK max & J = 14.50 $\pm$ 0.06;  +31.4 (days) & 5 \\
respect B max       & H = 13.74 $\pm$ 0.02;  +23.1 (days) & 5 \\
                    & K = 13.52 $\pm$ 0.02;  +22.3 (days) & 5 \\
$\Delta m_{15}(B)_{obs}$ & 1.12 $\pm$ 0.04  & 5 \\
$\Delta m_{15}(B)_{intrinsic}$& 1.25 $\pm$ 0.05  & 5 \\
$t_{r}$(B)          & 19.9 $\pm$ 0.5 & 5 \\
stretch factor in B & 0.97 $\pm$ 0.02 & 5 \\
$\Delta C_{12}$    & 0.39 $\pm$ 0.11 & 5 \\
Absolute magnitude & $M_U^{max}$ = -19.50 $\pm$ 0.94 & 5 \\
                   & $M_B^{max}$ = -19.17 $\pm$ 0.75 & 5 \\
                   & $M_V^{max}$ = -19.00 $\pm$ 0.49 & 5 \\
                   & $M_R^{max}$ = -18.78 $\pm$ 0.35 & 5 \\
                   & $M_I^{max}$ = -18.29 $\pm$ 0.19 & 5 \\
                   & $M_J^{max}$ = -18.22 $\pm$ 0.14 & 5 \\
                   & $M_H^{max}$ = -17.92 $\pm$ 0.11 & 5 \\
                   & $M_{K}^{max}$ = -18.08 $\pm$ 0.06 & 5 \\
log$_{10}L$        & 43.01 $\pm$ 0.05 \ergs & 5 \\
E(U-B)             & 1.08 $\pm$ 0.13 & 5 \\
E(V-R)             & 0.64 $\pm$ 0.06 & 5 \\
E(R-I)             & 0.74 $\pm$ 0.11 & 5 \\
E(V-J)             & 2.03 $\pm$ 0.20 & 5 \\
E(V-H)             & 2.34 $\pm$ 0.09 & 5 \\
E(V-K)             & 2.45 $\pm$ 0.23 & 5 \\
\hline
\end{tabular}
\begin{flushleft}
(1) NED; (2) \cite{schlegel98}; (3) Assuming a Virgo distance of
15.3 Mpc \citep{freedman01} and a relative distance from
Virgo of 1.18 for NGC 3169 (\citealt{kraan86}); (4) \cite{itagati03}; (5) This work. \\
\end{flushleft}
\end{table}

\section{Spectroscopy} \label{spec_results}
As can be seen in Figure \ref{fig_evolopt_spec}, a good set of
optical spectra were acquired for SN 2003cg, spanning -8.5 to
+52.5 days, plus a nebular spectrum taken 384.6 days after B
maximum. We also obtained ten early-time NIR spectra (Figure
\ref{fig_evolir_spec}) with NTT and UKIRT. This data set
constitutes a remarkable sequence of spectra and allows a detailed
IR spectroscopic study.

\subsection{Optical spectra} \label{opt_spec}
\begin{figure*}
\psfig{figure=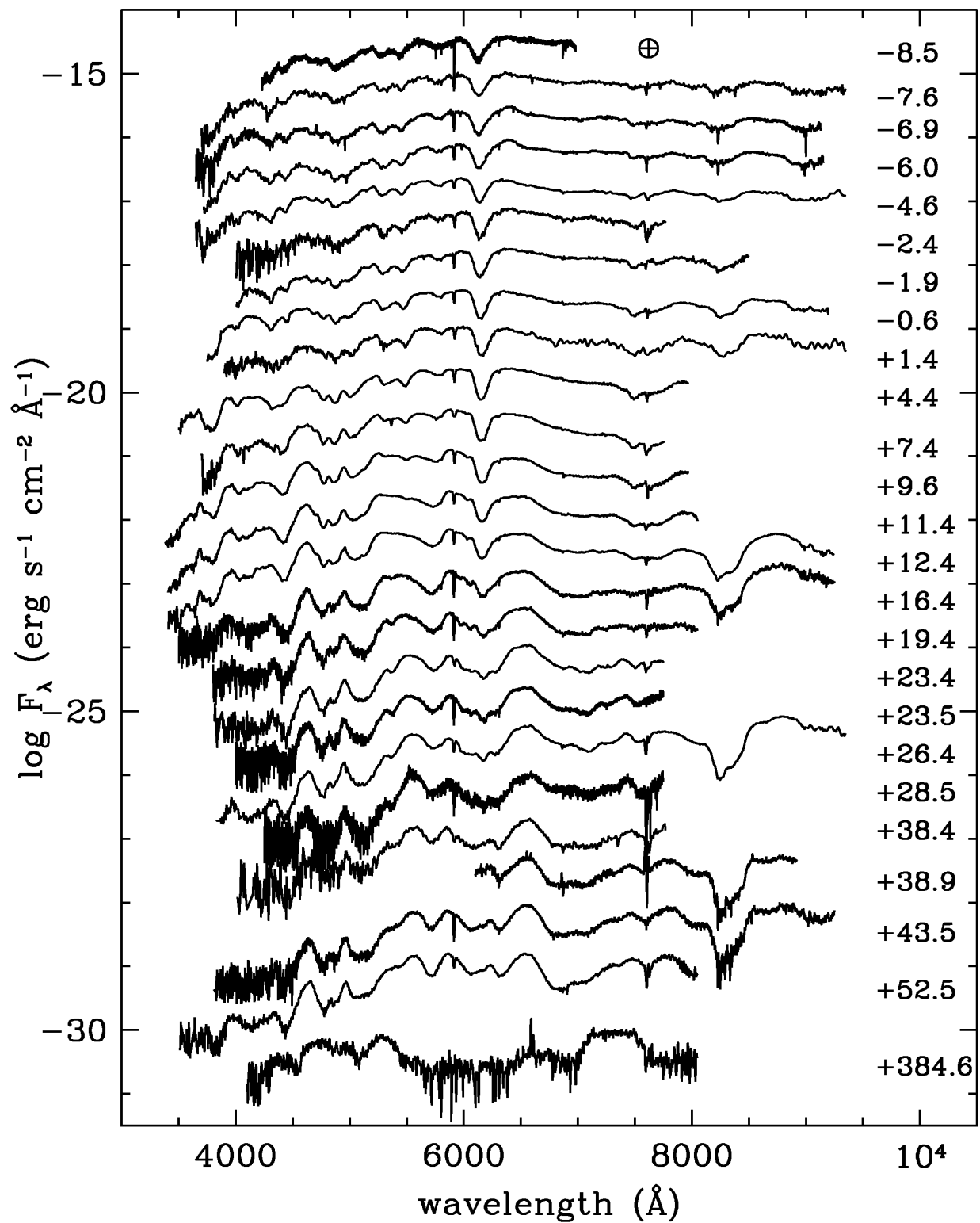,width=\textwidth,height=23cm}
\caption{Optical spectral evolution of SN 2003cg. The ordinate
refers to the first spectrum and the others have been shifted
downwards by arbitrary amounts. Epochs are given at the right hand
side. The $\oplus$ indicates the location of the strongest
telluric band, which has been removed when
possible.}\label{fig_evolopt_spec}
\end{figure*}

Figure \ref{fig_evolopt_spec} shows the optical spectral evolution
of SN 2003cg. It is similar to that of other 'normal' Type Ia SNe
apart from the red gradient due to the high reddening. Evidence of
high extinction towards SN 2003cg is apparent in all the spectra
in the form of strong (EW = 5.27 $\pm$ 0.50 $\AA$) \NaI~D
interstellar absorption at $\lambda \sim$ 5915 $\AA$. Some of the
spectra like that at -7.6 day also show interstellar \CaII\,H\&K
absorption at $\lambda$ 3934 and 3968 $\AA$. Around maximum light
the spectra show the P Cygni profile of \SiII~at $\sim$ 4130
$\AA$. Also two prominent absorptions are present in SN~2003cg at
$\sim$ 4300 $\AA$ and at $\sim$ 4481 $\AA$. These are attributed
to a combination of \MgII, \FeIII~and \SiIII~lines. Feature due to
\FeIII~and \SiIII~are again present around 5000 $\AA$. The
W-feature at $\sim$ 5300 $\AA$ is caused by \SII. Between 5960 and
6350 $\AA$ the spectra are dominated by \SiII~lines. The earliest
\CaII~IR triplet spectra include a weak high velocity component.
For the -8.5 day spectrum we find \textit{v} $\sim$ 22000 \kms for
this component. The high-velocity feature disappears after maximum
light. \cite{mazzali05} carried out a detailed study of the high
velocity feature in a number of SNe~Ia including SN~2003cg.\\

A single late-phase (nebular) spectrum was secured with the ESO
VLT-UT2+FORS1 at +384.5 days. A spectrum at this late phase is
particularly valuable since the ejecta are mostly transparent to
optical/NIR light. Thus, via modelling it is possible to estimate
directly the total mass of $^{56}$Ni synthesised in the explosion.
Mazzali et al. (in preparation) estimate $M_{Ni}$ =
0.53 $\pm$ 0.05 \M~for this SN~Ia.\\

\begin{figure*}
\psfig{figure=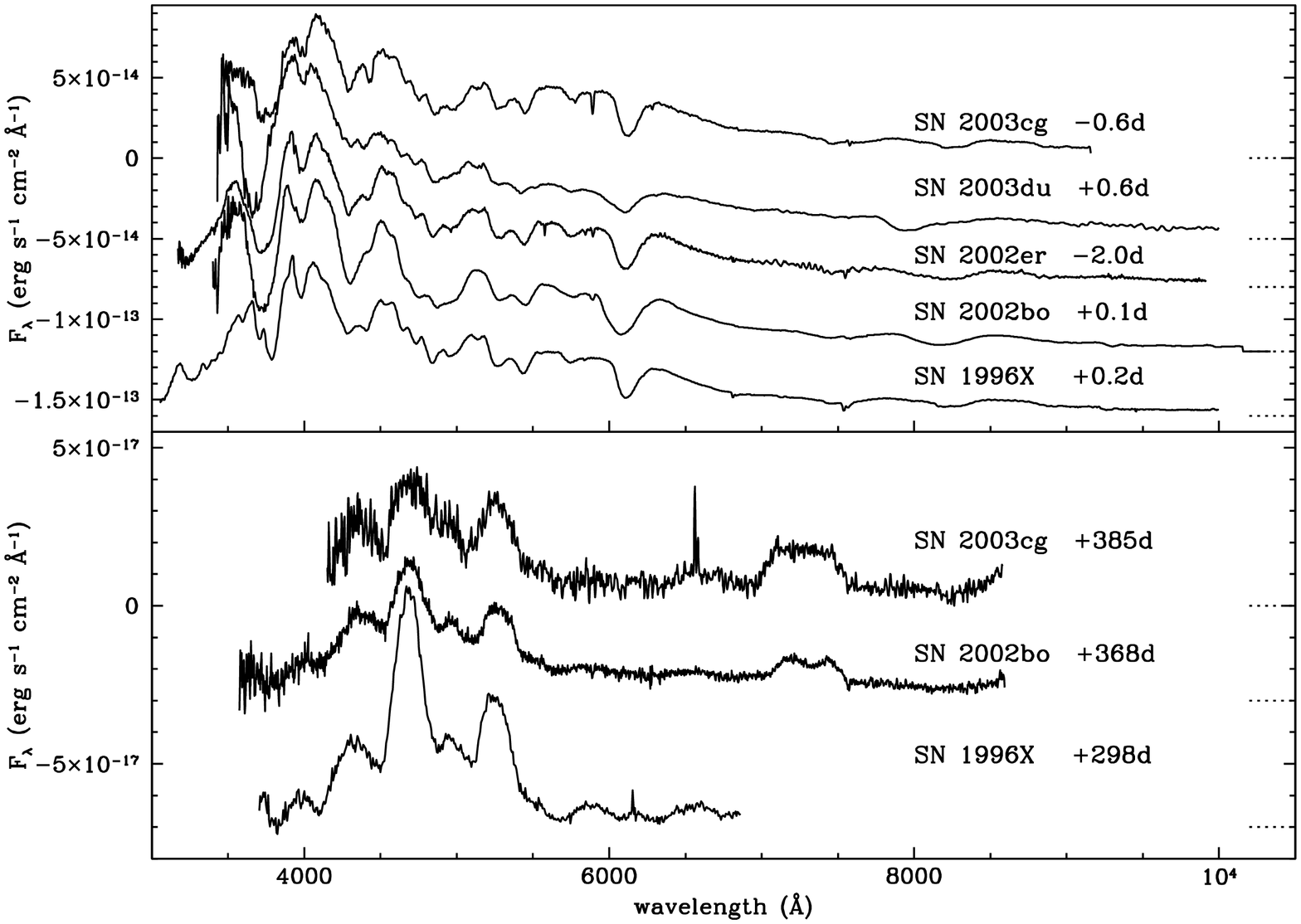,angle=270,width=\textwidth}
\caption{Comparison between optical spectra of SN 2003cg with
those of SNe 2003du (Stanishev et al. in preparation), 2002er
(\citealt{kotak05}), 2002bo (\citealt{benetti04}) and 1996X
(\citealt{salvo01}) at maximum light (top) and at late-phase
(bottom). The spectra have been displaced vertically for clarity
(the zero-flux levels are indicated by the dotted lines). All the
spectra have been corrected for redshift and reddening (see
text).}\label{fig_comopt_spec}
\end{figure*}

In Figure \ref{fig_comopt_spec} (top) we compare the SN 2003cg
optical spectra around maximum light with those of other SNe Ia
such as SNe~2003du (Stanishev et al. in preparation), 2002er
(\citealt{kotak05}), 2002bo (\citealt{benetti04}) and 1996X
(\citealt{salvo01}). Comparison of the SN~2003cg nebular phase
spectrum with those of SNe~2002bo and 1996X is also shown in
Figure \ref{fig_comopt_spec} (bottom). The $\Delta m_{15}$ values
of these SNe~Ia are in the range 1.07-1.33, close to the average
for SNe~Ia. In spite of this relative homogeneity there are some
clear differences between the spectra. At maximum light both SN
2003cg and SN~2002er show well-developed, structured features due
to intermediate mass elements but with clear differences in the P
Cygni profiles up to 4500 $\AA$. However, the same features in SN
2003du are less pronounced, possibly due to high expansion
velocity or composition effects. The SN~Ia-defining
\SiII~$\lambda$ 6355 line is clearly visible until $\sim$ 28 days
after maximum. The line has a fairly symmetrical profile as is
also the case for SNe~1996X and 2002er. Somewhat less symmetrical
profiles are seen in SNe 2002bo and 2003du.  The \SiII~($\lambda$
6355 $\AA$) trough blueshifts are similar for SN 2003cg and SN
2003du (see Section \ref{veloc}). However the pre-maximum light
\CaII\,H\&K ($\lambda$ 3934 and 3968 $\AA$) trough blueshifts in
SN~2003du are higher by a factor $\sim\times$1.2
and decline more slowly with time than in SN 2003cg.\\

In the nebular spectrum of SN 2003cg (Figure \ref{fig_comopt_spec}
- bottom) the [\FeII]/[\FeIII] emission feature at $\sim$
4600~$\AA$ is comparable to that seen in SN~2002bo but is less
pronounced than in SN~1996X. For this feature, the position of
SN~2003cg on a plot of FWHM vs. \dm15 coincides with the average
regression curve for SNe~Ia. (\citealt{mazzali98b}- Figure~2). The
[\CoII] 5900 $\AA$ emission \citep{axelrod80} appears barely
visible above the noise. The boxy feature between 7000 and 7600
$\AA$ may be due to a blend of [\FeII]
and [\CaII] (Figure \ref{fig_comopt_spec} - bottom). \\

\subsection{Near-infrared spectra} \label{ir_spec}
\begin{figure*}
\begin{center}
\psfig{figure=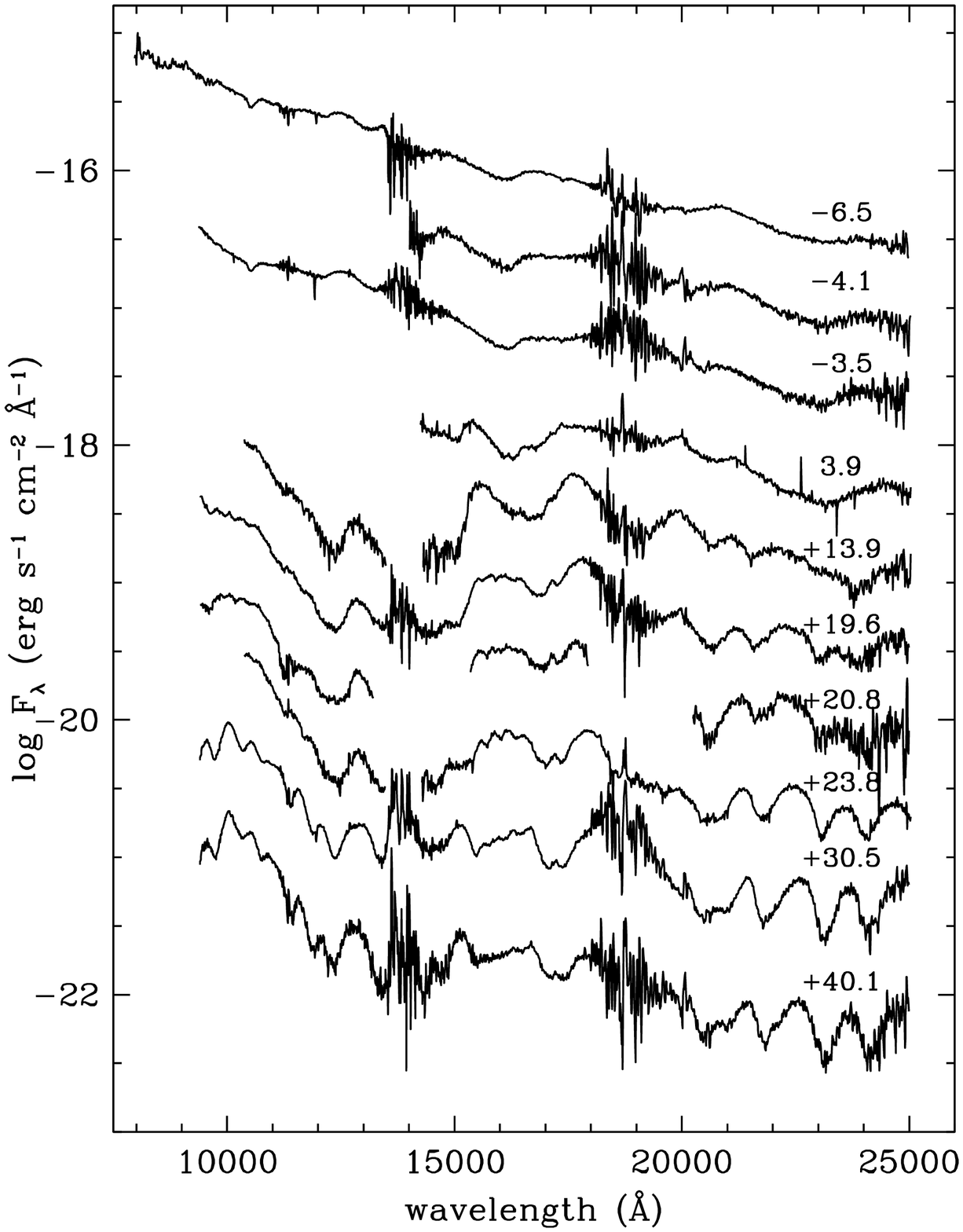,width=\textwidth,height=23cm}
\caption{NIR spectral evolution of SN 2003cg. The ordinate refers
to the first spectrum and the others have been shifted downwards
by arbitrary amounts. Epochs are shown to the right.}
\label{fig_evolir_spec}
\end{center}
\end{figure*}

\begin{figure*}
\begin{center}
\psfig{figure=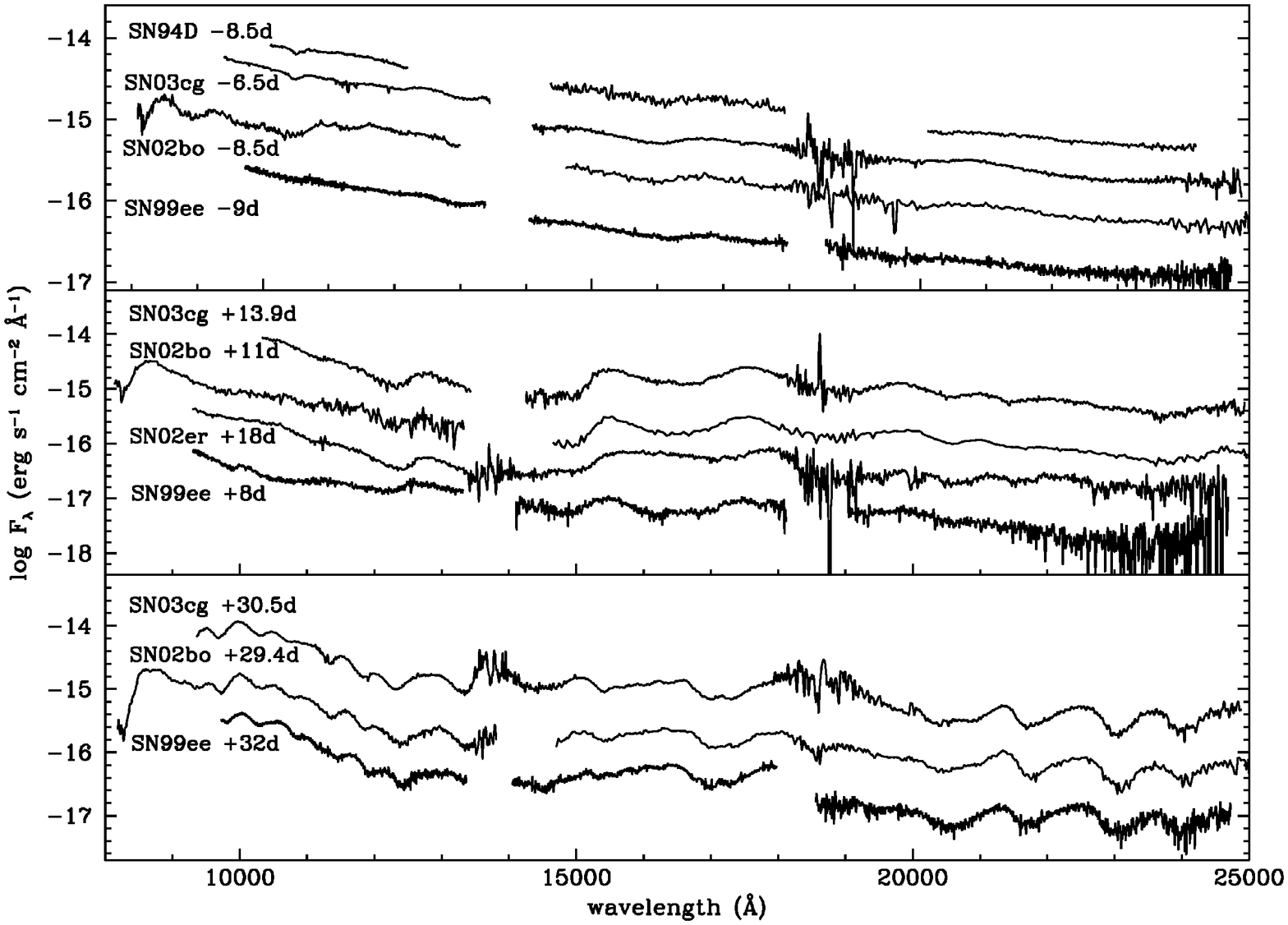,angle=270,width=\textwidth}
\caption{Comparison between the NIR spectra of SN 2003cg and those
of SNe 1994D (\citealt{meikle96}), 1999ee (\citealt{hamuy02}),
2002bo (\citealt{benetti04}) and 2002er (\citealt{kotak05}) before
maximum (top), two weeks after maximum (middle) and one month
after maximum (bottom). The spectra have been displaced vertically
for clarity. All spectra have been corrected for reddening (see
text) and parent galaxy redshift.} \label{fig_comir_spec}
\end{center}
\end{figure*}

Near-infrared (NIR) spectra are important because there is reduced
line blending and lower line opacity.  Consequently, line
identification is less ambiguous and we may look more deeply into the
ejecta. Thus, the line intensities and profiles provide a tool for
analysing the ejecta composition and stratification.  In Figure
\ref{fig_evolir_spec} we show the NIR spectral evolution of SN
2003cg. The spectral evolution is typical of SNe~Ia. The earliest
spectra are dominated by continuum emission but, as the photosphere
recedes, the spectra are increasingly dominated by discrete,
doppler-broadened features. In Figure \ref{fig_comir_spec} we compare
three early-phase (--6d, +14d, +31d) NIR spectra of SN 2003cg with
approximately coeval spectra of SNe 1994D (\citealt{meikle96}), 1999ee
(\citealt{hamuy02}), 2002bo (\citealt{benetti04}) and 2002er
(\citealt{kotak05}).\\

In the pre-maximum spectra of SN~2003cg a weak P Cygni profile is
visible at about 10500~$\AA$. This feature was first noted in
pre-maximum spectra of SN 1994D by \cite{meikle96}. This was
discussed by \cite{meikle96,mazzali98,hamuy02,benetti04} and
attributed to \HeI~10830 $\AA$ or \MgII~10926 $\AA$.  On the basis
of their modelling, \cite{meikle96} and \cite{mazzali98} concluded
that \HeI, \MgII~or a mixture of the two could be responsible for
the feature. In contrast, according to the NIR spectral models of
\cite{wheeler98}, the feature must be due almost entirely to
\MgII~as there is not enough He in the model atmosphere to form a
He line.  However, simultaneous optical-IR spectral model fits to
SN 1994D by \cite{hatano99} and examination of the -7 day spectrum
of SN 2002cx by \cite{nomoto03} has led these authors to challenge
the \MgII~identification.  They find that the \MgII~interpretation
also predicts non-existent features at other wavelengths. It is
possible that the \MgII~explanation is only appropriate for faster
decliners. Regardless of which identification we adopt, we find
that the trough blueshift of SN~2003cg at --6.5~days is similar to
that seen in SN
1994D at -8.5 days.\\

In the SN 2003cg and SN 1999ee spectra \FeIII~12550 $\AA$ emission is
visible (Figure \ref{fig_comir_spec}). In addition, a broad P
Cygni-like feature is present at $\sim$ 16700 $\AA$. According to
\cite{wheeler98}, this is due to \SiII. In the $K$-band, emission at
20500~$\AA$ is visible in SN 2003cg. This was also seen in SN~2002bo,
albeit more weakly, and was identified with \SiIII~\citep{benetti04}.

By +1 to +2 weeks, the weak \MgII/\HeI~$\sim$10800 $\AA$
absorption has disappeared and strong absorption/emission features
have appeared. The prominent emission at 13000~$\AA$ is attributed
to \SiII, while the strong features around 15500-17500 $\AA$ are
produced by iron group elements such as \CoII, \FeII~and
\NiII~(\citealt{wheeler98}). In the $K$~band \SiII~and Fe-group
lines have appeared (\citealt{wheeler98}). This indicates that, by
this epoch, the NIR photosphere has already receded through the
intermediate mass element zones and has penetrated quite deeply
into the Fe-group region.  In this respect, SN 2003cg is more
similar to SN 2002bo than to SN 2002er whose spectrum was more
characterised by lighter elements during this era. Like SN~1999ee,
SN~2003cg presents spectral features due to Mg, Ca and the
Fe-group elements around 12000 $\AA$ but at a lower strength.\\

By one month post-maximum light the three broad emission peaks in
the range 21350-22490 $\AA$, produced by Co, Ni and Si, have
become more prominent. This is similar to the behaviour seen in
other SNe~Ia
over this spectral range and phase.\\

\begin{figure*}
\psfig{figure=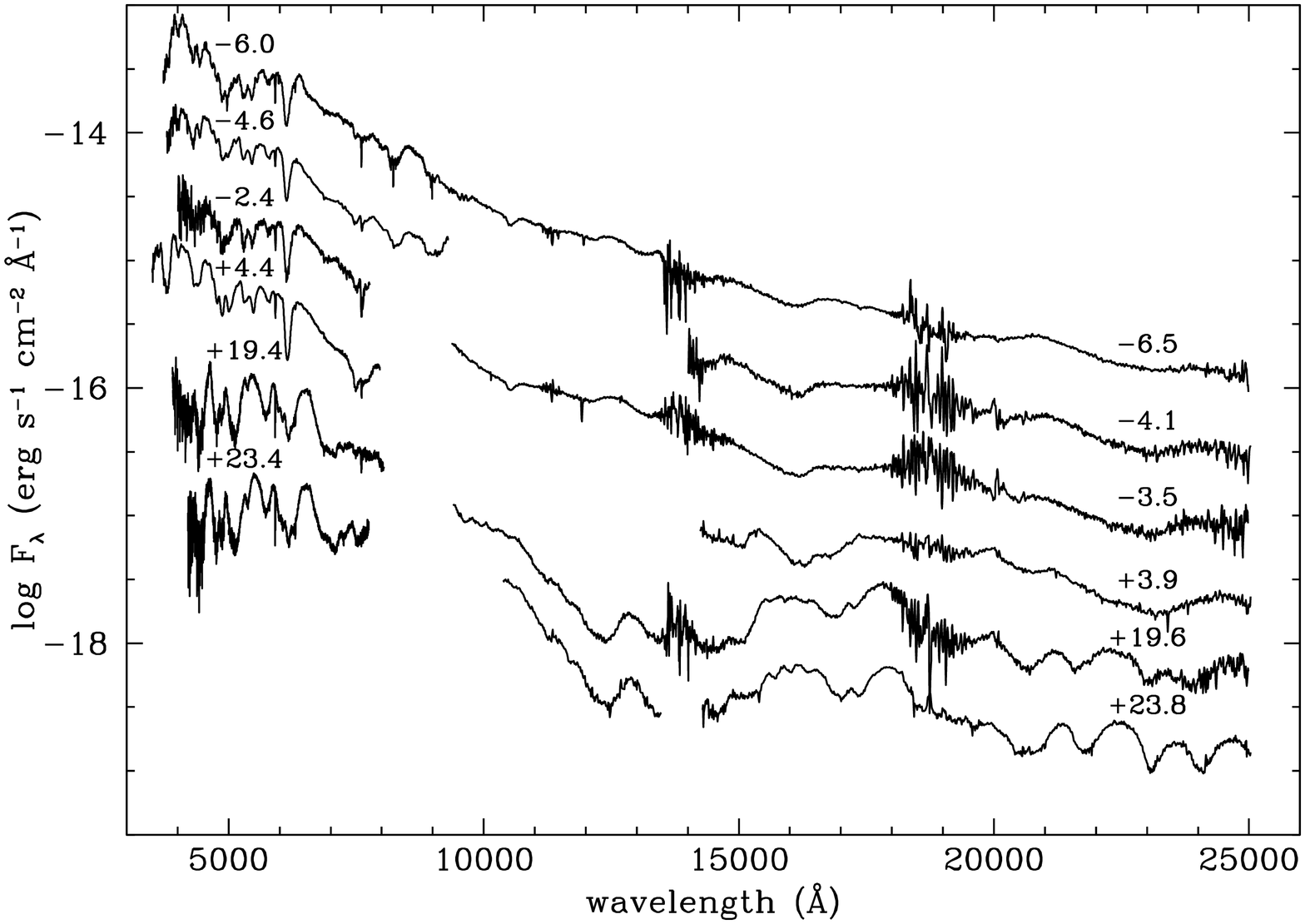,angle=270,width=\textwidth}
\caption{Combined optical and IR spectra, showing the evolution of SN
2003cg from 1 week before to 3 weeks after maximum light. The spectra
have been corrected for reddening due to the Milky Way and the host
galaxy.}
\label{fig_optirunion_spec}
\end{figure*}

Figure \ref{fig_optirunion_spec} shows combined coeval optical and
IR spectra of SN 2003cg, allowing a more complete view of the SED
evolution. We find that, at -6.5 days the NIR flux (integrated
between 10000 and 25000 $\AA$) contribution was about 8$\%$ of the
total flux, rising to about 14$\%$ at +23.8~days. For the same
interval, \citet{suntzeff96,suntzeff03} estimated the contribution
of IR in SNe Ia to be approximately 15$\%$ of the total flux.

\subsection{Velocity Gradient} \label{veloc}
\begin{figure}
\psfig{figure=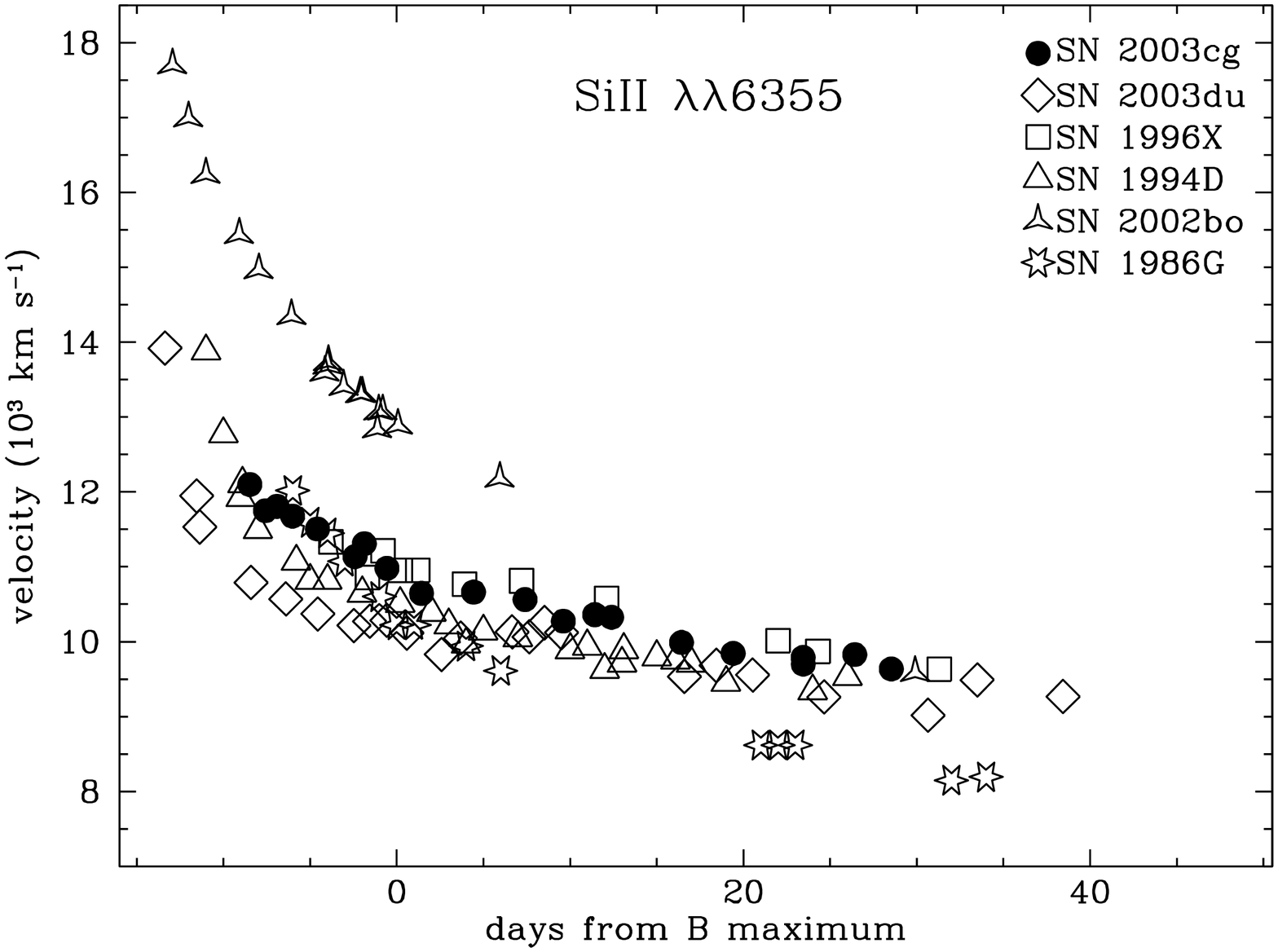,angle=270,width=9cm,height=7.5cm}
\caption{Evolution of the expansion velocity derived from the
minima of \SiII~6355 $\AA$ for SN 2003cg, compared with other LVG
SNe: SN 2003du (Stanishev et al., 2004, in preparation), SN 1996X
(\citealt{salvo01}) and SN 1994D (\citealt{patat96}). Also shown
are the evolution of the HVG SN 2002bo (\citealt{benetti04}) and
the FAINT SN 1986G (\citealt{cristiani92}, \citealt{benetti05}).}
\label{fig_vel_si2}
\end{figure}

As can be seen in Figure \ref{fig_evolopt_spec}, SN 2003cg
exhibits a deep and symmetric \SiII~trough at 6355 $\AA$, visible
approximately from --8 to +28 days, and shifting rapidly redwards
with time. In Figure \ref{fig_vel_si2}, we compare the SN~2003cg
\SiII~velocity evolution with those of a sample of SNe~Ia.
Clearly, SN~2003cg lies within the low velocity gradient group
(hereafter LVG)\footnote{\cite{benetti05} have analysed the
photometric and spectroscopic diversity of 26 SNe~Ia. They
identify three groups: FAINT (faint SNe Ia with low expansion
velocities and rapid evolution of the \SiII~velocity), HVG
('normal' SNe Ia with high velocity gradients, brighter absolute
magnitudes and higher expansion velocities than the FAINT SNe) and
LVG ('normal' and SN 1991T-like SNe Ia with small velocity
gradients).}. The SN~2003cg velocity gradient \textit{\.{v}}
$\sim$ 38 $\pm$ 6 \kms $d^{-1}$ and its expansion velocity at 10
days past maximum $\textit{v}_{10}$(\SiII) = 10310 $\pm$ 0.10
\kms. These are typical LVG values. By way of comparison, we also
show in Figure \ref{fig_vel_si2} the velocity evolutions of
the HVG SN 2002bo (HVG) and the FAINT SN 1986G.  \\

\begin{figure}
\psfig{figure=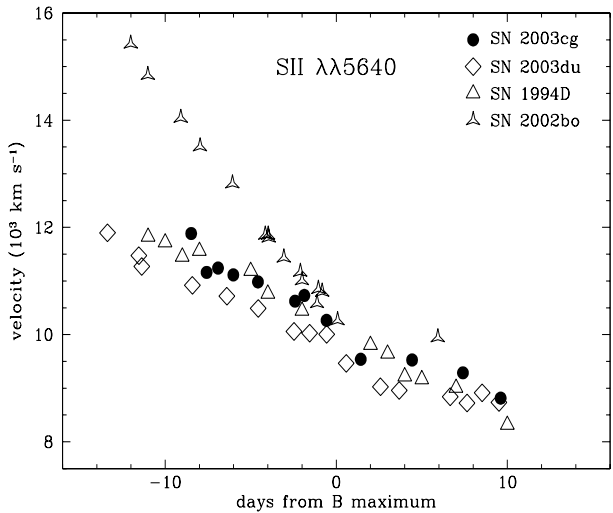,angle=270,width=9cm,height=7.5cm}
\caption{Evolution of the expansion velocity derived from the
minima of \SII~5640 $\AA$ for SN 2003cg, and two other LVG SNe: SN
2003du, SN 1994D. Also shown is the evolution of the HVG SN
2002bo.} \label{fig_vel_s2}
\end{figure}

In Figure \ref{fig_vel_s2} we show the velocity evolution of the
\SII~5640 $\AA$ trough blueshift, compared with those of three
other normal SNe~Ia. This feature tends to disappear more quickly
than does \SiII, and is generally difficult to discern beyond
+10~days. SNe~Ia show two types of behaviour here, exhibiting
either a steep velocity gradient (e.g. SN 2002bo) or a shallow
gradient (e.g. SNe 1994D, 2003du) \citep{benetti05b}. It
may be seen that SN~2003cg falls into the latter category.  \\

\begin{figure}
\begin{center}
\psfig{figure=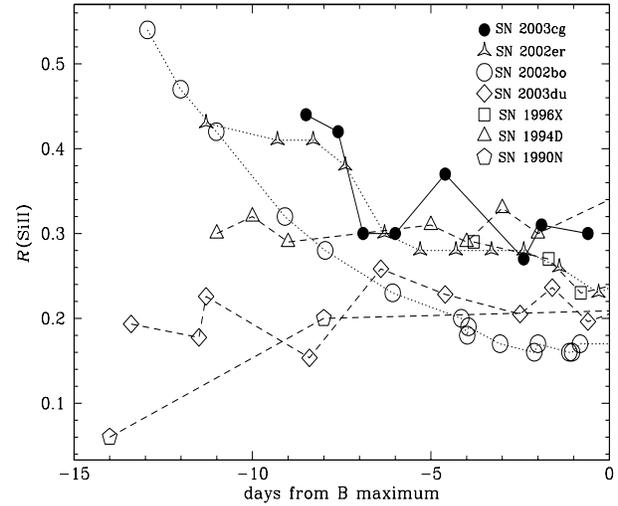,angle=270,width=8.8cm,height=7.5cm}
\caption{Pre-maximum evolution of $\cal R$(\SiII) for SN 2003cg,
compared with those SNe 2002er, 2002bo (HVG) 2003du, 1996X, 1994D
and 1990N (LVG) \citep{benetti05}.}\label{fig_evorsi}
\end{center}
\end{figure}

$\cal R$(\SiII), the ratio of the depths of the \SiII~5972 $\AA$
and 6355 $\AA$ absorptions, was introduced by \cite{nugent95} as a
potential distance-independent parameter which is related to the
luminosity. The parameter was investigated further by
\cite{benetti05}. For SN~2003cg, $\cal R$(\SiII$)_{max}$ = 0.30
$\pm$ 0.05.  In Figure \ref{fig_evorsi} it may be seen that the
pre-maximum evolution of $\cal R$(\SiII) for SN 2003cg is similar
to that of other LVG SNe~Ia i.e. little or no pre-maximum
evolution
occurs.\\

\begin{figure}
\psfig{figure=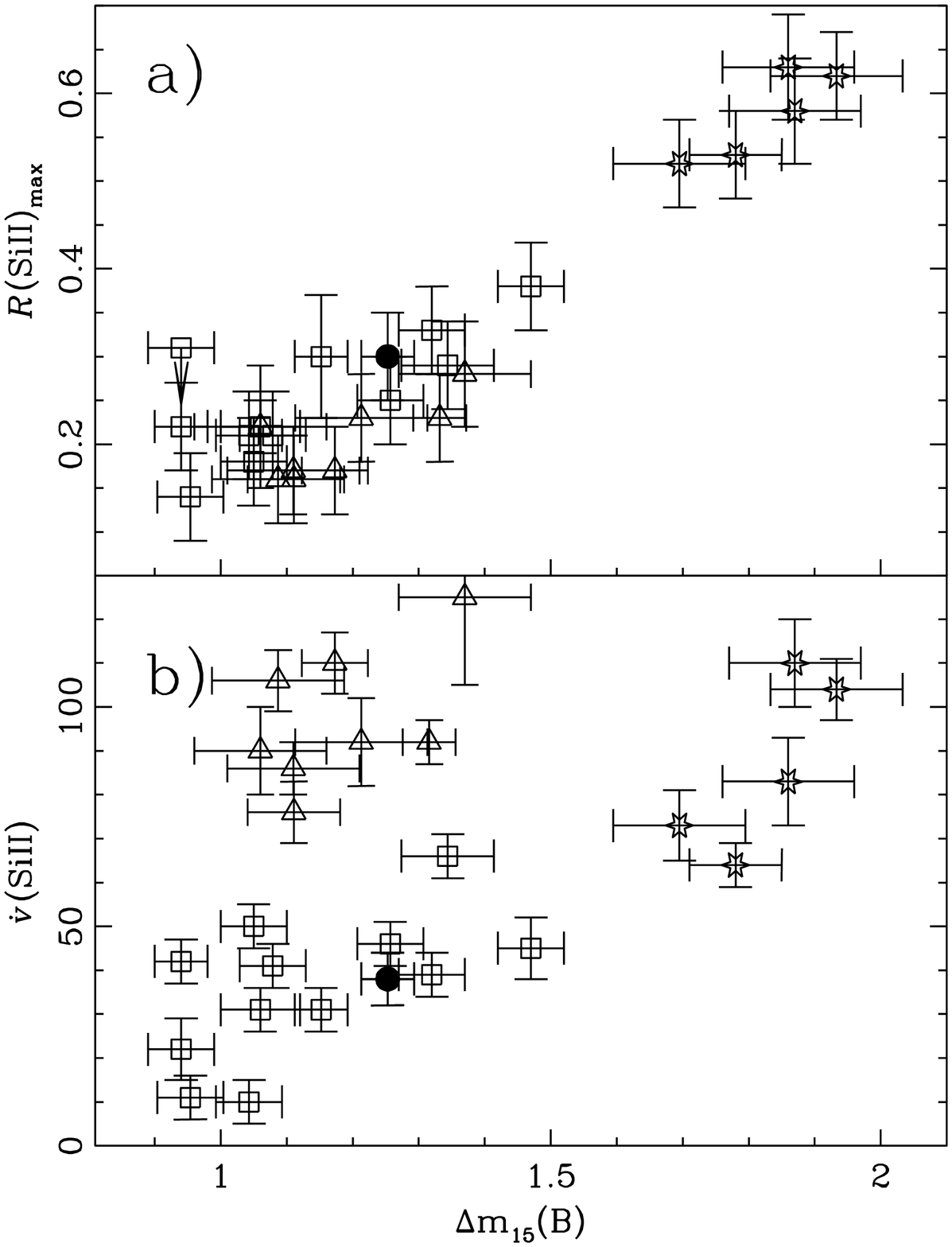,width=8cm,height=9cm}
\caption{a) $\cal R$(\SiII)$_{max}$ vs. \dm15. b) \textit{\.{v}}
vs. \dm15 for \SiII $\lambda$ 6355. Filled symbols refer to SN
2003cg, and squares, triangles and stars represent LVG, HVG and
FAINT SNe, respectively \citep{benetti05}.}
\label{fig_rsi_vel_dm15}
\end{figure}

We repeated the \cite{benetti05} cluster analysis including SN
2003cg.  We found that this SN does indeed belongs to the LVG
cluster (see Figure \ref{fig_rsi_vel_dm15}). In the five-parameter
space (\dm15, $M_B^{max}$, \textit{\.{v}},
$\textit{v}_{10}$(\SiII) and $\cal R$(\SiII$)_{max}$), the nearest
SNe to SN 2003cg are SN 1994D and SN 1996X.  This supports our
choice of these SNe for comparison purposes in this work.

\subsection{Spectral modelling} \label{models}
We have compared the observed spectra of SN~2003cg with synthetic
spectra derived using the Lucy-Mazzali Monte Carlo code, which has
been used successfully to model spectra of SNe Ia in the
photospheric phase (e.g. \citealt{mazzali93}). The code
\citep{mazzali93,lucy99,mazzali00} assumes that the SN ejecta can
be separated by a sharp photosphere into an optically thick region
below which all the light is emitted and an optically thin region
where line formation occurs. The code's input parameters are the
SN luminosity, a photospheric velocity (which is equivalent to
radius since $v = r/t$, a density structure and a set of
abundances. The density structure used here is that of the W7
model \citep{nomoto84}. The flux at the photosphere is assumed to
be emitted with a black-body spectrum. The code follows the
propagation of photons in the ejecta and their interaction with
lines (including the process of branching) and electron
scattering. Excitation and ionisation are computed using a nebular
approximation, which gives a good description of the conditions in
a SN Ia near maximum \citep{pauldrach96}. The emerging spectrum is
computed using a formal integral.
\subsubsection{Days --7.6/--6.5}
\begin{figure}
\centering
\psfig{figure=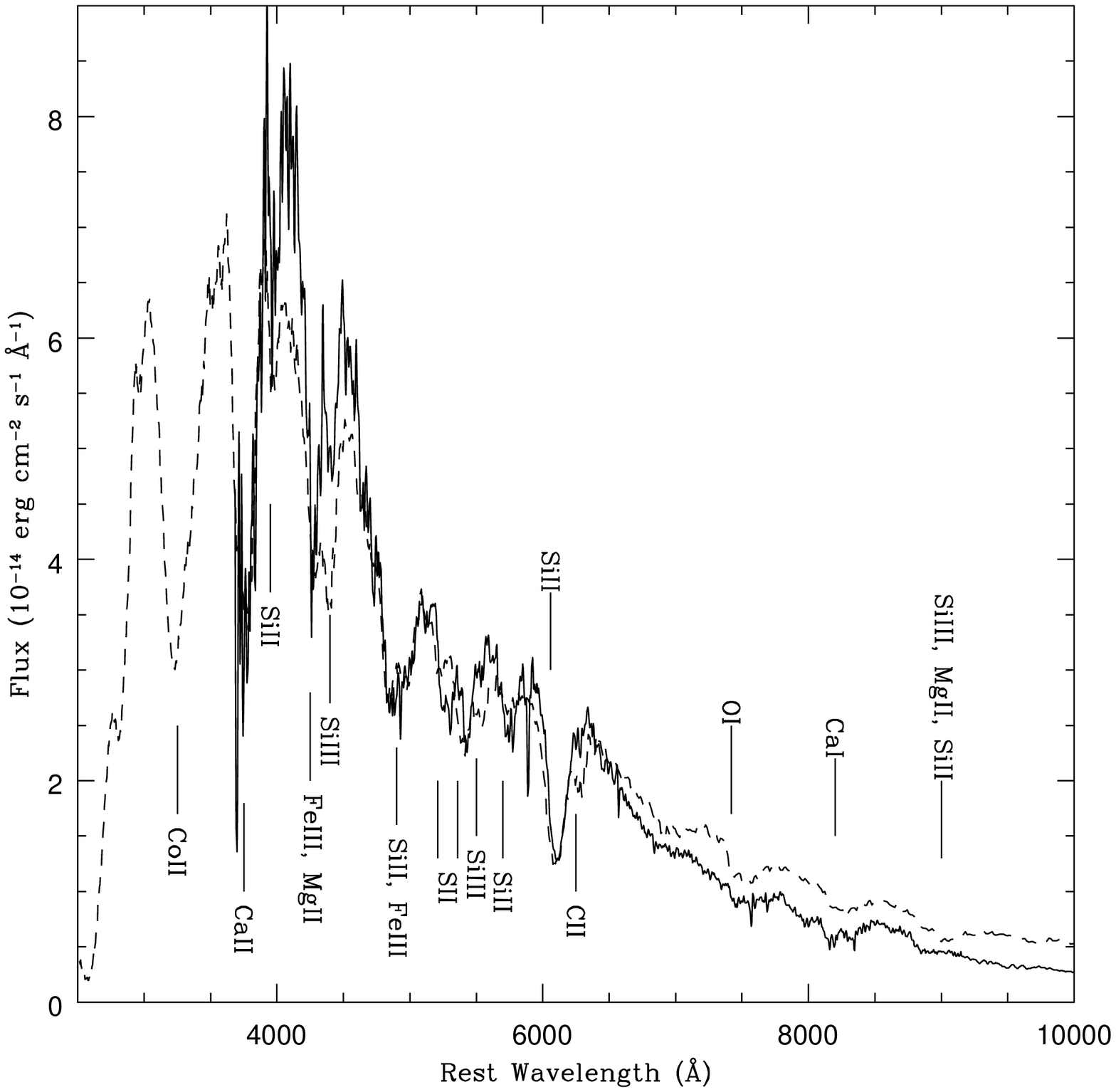,width=8.5cm,height=6.4cm}
\psfig{figure=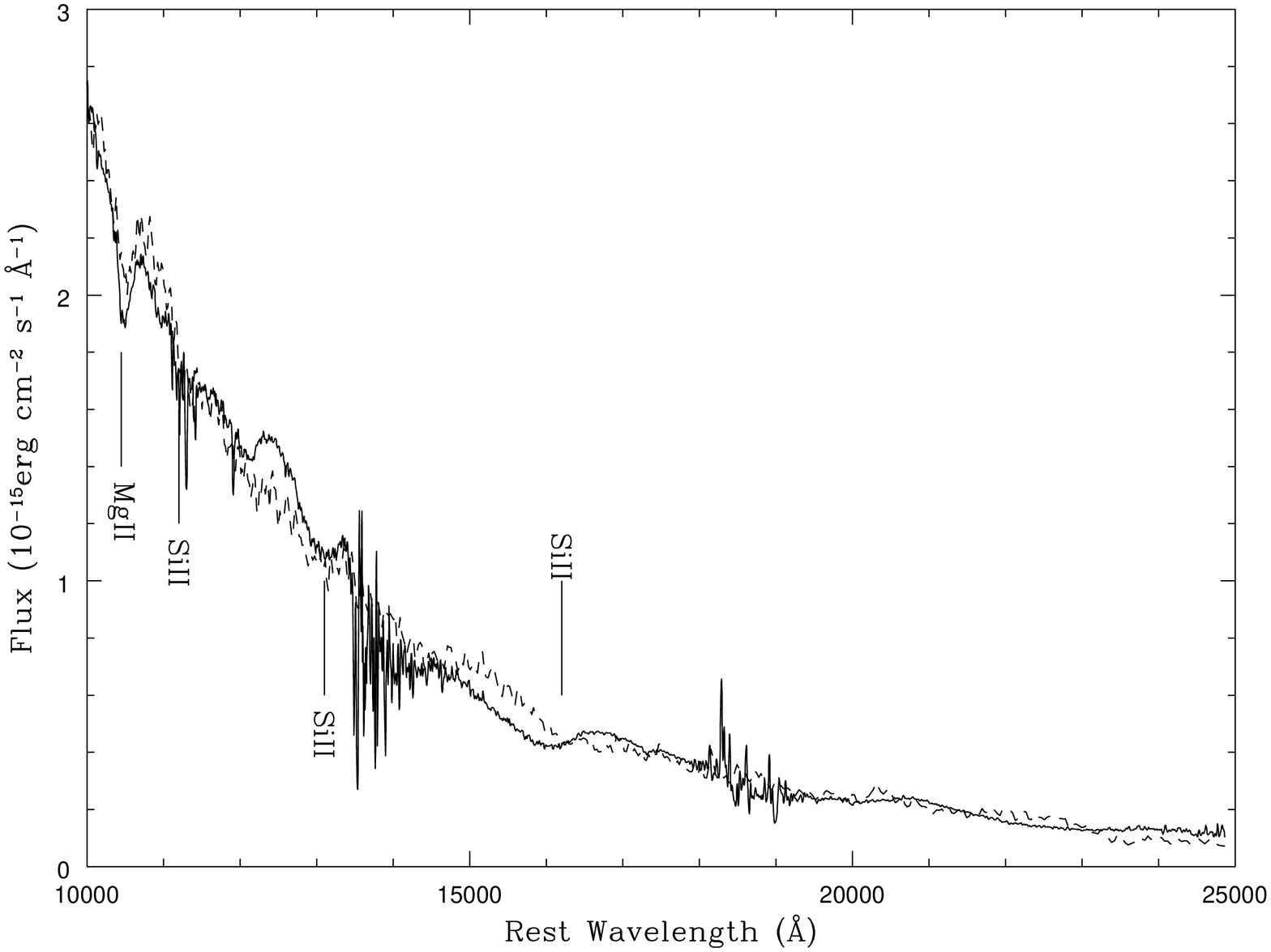,angle=270,width=8.5cm,height=6.4cm}
\caption{Observed, dereddened optical (top) and NIR (bottom)
spectra of SN~2003cg at day $-7.6$/$-6.5$ and corresponding
spectral models. The models parameters include log$_{10}L =
42.88$\ergs, $v_{ph} = 10,300$\,\kms, $\mu = 31.28$ and $t_{exp} =
12.3$\,d.} \label{min7_6_der}
\end{figure}

Here we model the day~$-7.6$ spectrum optical spectrum and
day~$-6.5$ NIR spectrum, which were among the earliest obtained.
In the model, the SN ejecta above the photosphere is divided into
three spherically symmetric shells above 11200, 11600, and
15500\,\kms, respectively. All shells are dominated by O (48\% to
64\% by mass). The outermost shell contains 5\% C to account for
the \CII\ line (see below). A rather high amount of Si (18\% to
25\% in the inner shell) is necessary to reproduce the deep \SiII\
features in the spectrum. Also of interest is the high Fe~group
abundance at this early stage. We find 5\% Ni in all shells. The
stable Fe abundance, i.e. Fe that is not produced in the Ni decay
chain, is between 1.5\% near the photosphere, and 0.5\% in the
outermost shell (for reference the solar mass fraction of Fe is
$\approx 0.27\%$ - \citealt{gratton03}). Finally, Ti and Cr of
$\approx 0.25\%$ are needed in order to shift the flux from the UV
to optical wavelengths (the solar mass fraction is $\approx
6.22\times10^{-4}\%$ for Ti and $\approx 3.80\times10^{-3}\%$ for
Cr - \citealt{gratton03}). The best fit was achieved using
log$_{10}L = 42.88$\ergs, a photospheric velocity $v_{ph} =
10,300$\,\kms, and epoch after explosion $t_{exp} = 12.3$\,d (i.e.
a rise time of 19.9~days, cf. Section \ref{earl_ph}). The observed
spectrum, dereddened according to the prescription derived above
(Section \ref{indredd}), and the synthetic spectrum are compared
in Figure \ref{min7_6_der} (top), where the main
features are also identified.\\
The model reproduces the overall shape and the individual line
profiles reasonably well, but it has difficulties matching the
very blue $B-V$ colour obtained after dereddening. Achieving a
bluer spectrum may require increasing the metal abundance near the
photosphere to increase line blocking and therefore transfer more
near-UV photons to the B region. given the lack of info on the U
band flux, we cannot constrain this aspect very tightly. In
particular, in order to obtain a blue continuum we have to adopt a
rather low value for the photospheric velocity. This leads to
near-photospheric temperatures that are too high, and results in
the presence of lines in the synthetic spectrum that are not
present in the observed one. In particular this affects \SiIII\
lines.

Starting at the blue end of the spectrum, the strong \CaII\,H\&K
near 3800\AA\ can be seen. The narrow feature at $\approx
4000$\AA\ is identified as \SiII\ 4130\AA. The peak near 4000\AA\
is suppressed in the model due to \SIII\ absorptions. This
indicates that the temperature near the photosphere is too high.
The excessive strength of two other \SiIII\ lines, 4565\AA, which
is seen in the model near 4400\AA, and 5740\AA, seen near 5500\AA,
substantiates this assumption. The absorption at 4250\AA\ is a
combination of \MgII\ 4481\AA, \FeIII\ 4420\AA, and \SiIII\
4339\AA. The feature slightly blueward of 5000\AA\ is caused by
\SiII\ 5056\AA\ and \FeIII\ 5128, 5156\AA. In this case the
presence of strong \FeIII\ lines is supported by the accuracy with
which the observed profile is matched. Near 5300\AA, we recognize
the typical \SII\ W-feature, followed by the two \SiII\ lines at
5960\AA\ and 6350\AA. The weakness of the synthetic \SiII\
5960\AA\ line again suggests that the model temperature near the
photosphere is too high. The emission component of the \SiII\
6350\AA\ line is suppressed in the observed spectrum. This effect
is reproduced in the model by \CII\ 6580\AA\ absorption. This has
also been detected in other SNe~Ia (e.g. \citealt{mazzali01}).
Further to the red only two more lines are clearly identified:
\OI\ 7774\AA, and the \CaII~IR triplet. Both lines are somewhat
deeper in the model. An absorption in the synthetic spectrum near
9000\AA\ is attributed to a blend of \MgII\ 9218\AA, \SiIII\
9324\AA, and \SiII\ 9413\AA.

In Figure \ref{min7_6_der} (bottom) we compare the day~$-6.5$ NIR
spectrum with the model spectrum described above, extended to the
NIR. Owing to the black body lower boundary adopted in the model,
the predicted IR continuum exceeds that observed. We attribute
this to a drop in opacity which appears redward of $\sim$7000\AA\
and is not reproduced in the model (see Figure \ref{min7_6_der} -
top). Therefore, to match the observed spectrum, the model
continuum was multiplied by a factor 0.5.  However, apart from the
discrepancy in the overall flux level, the shape of the IR
continuum and other features are well reproduced by the model at
this epoch.  Only a few, relatively weak lines are present in the
NIR spectrum, primarily due to \SiII\ and \MgII. The feature at
$1.05 \mu$ is well reproduced by \MgII\ 10914, 10951~\AA. As well
discuss \cite{hatano99} in their work (see also Section
\ref{ir_spec}), the identification of \MgII\ at this wavelength
implies the presence of other \MgII\ features in other parts of
the optical/NIR spectrum, some of which have been identified
above. In spite of this and due the lower strength of this lines
in the observed spectra, we can not exclude the presence of He
10830\AA.
\subsubsection{Day --0.6}
\begin{figure}
\centering
\psfig{figure=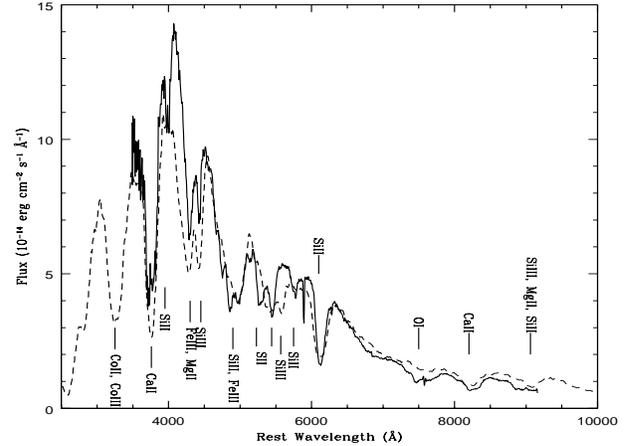,width=8.5cm,height=6.4cm}
\caption{Observed, dereddened optical spectrum of SN~2003cg at day
$-0.6$ and corresponding spectral model.  The model parameters
include log$_{10}L = 43.11$\ergs, $v_{ph} = 6,000$\,\kms, $\mu =
31.28$ and $t_{exp} = 19.3$\,d.} \label{min0_6_der}
\end{figure}

We have also modelled the SN~2003cg optical spectrum near maximum
light. The dereddened day~$-0.6$ spectrum was compared to a
synthetic  spectrum computed using the following  parameters:
$t_{exp} = 19.3$\,d, log$_{10}L = 43.11$\ergs\ and $v_{ph} =
6000$\,\kms. The observed spectrum, dereddened as above, and the
synthetic spectrum are compared in Figure \ref{min0_6_der}, where
the main features are also identified.\\

As for the pre-maximum spectrum, we are forced to use a very low
velocity for this epoch because the dereddened spectrum is
unusually blue. Our synthetic spectrum has $B-V \sim 0.0$, but the
dereddened spectrum has $B-V \sim -0.2$. Therefore, on the one
hand our model colour is too red, and on the other our model line
features are somewhat too hot compared to the observed ones.

The features present in the spectrum are generally similar to
those deduced in the --7.6~day spectrum. In the blue region
\CaII\,H\&K is strong. \SiII\ 4130\AA\ is present in the data but
not reproduced in the model. The line near 4300\AA~is now
dominated by \FeIII\ 4419, 4431\AA\ and \MgII\ 4481\AA. Moving
redwards, we again identify \SiIII\ 4560\AA. This line is strong
both in the observed spectrum and in the synthetic one. However,
other lines of \SiIII, notably 5740\AA, which is strong in the
model near 5500\AA, but are absent in the data, indicate that the
model temperature is too high. The feature near 5000\AA\ is
dominated by \FeIII\ 5127, 5156\AA, with contributions from \SiII\
5056\AA\ and \FeII\ 5169\AA. The observed profile of this feature
however suggests that \FeII\ lines also make an important
contribution. The \SII\ W-feature is affected by the presence of
\SiIII\ 5740\AA. The prominent \SiII\ 6350\AA\ line is well
reproduced with an increased Si abundance (reaching $\sim 40$\%
near the photosphere. The \SiII\ 5960\AA\ is however not
reproduced at all, again indicating that the adopted velocity and
luminosity lead to a temperature that is too high. By this epoch,
the \CII\ line on top of the \SiII\ 6350\AA\ emission has
vanished. This is because the density in the outermost layers,
where carbon is found, is now too low for the line to form.
Further to the red we identify strong \OI\ 7774\AA, and the
\CaII~IR triplet. Finally, we identify another strong absorption
at 9000\AA\ due mostly to \SiIII\ 9324\AA.

\section{Summary} \label{summary}

SN 2003cg is a heavily reddened but otherwise normal Type Ia
supernova. We have presented the results of an intensive
optical/NIR monitoring programme by the ESC using a wide range of
telescopes and instruments. Photometry and spectroscopy were
acquired spanning day -8 to +414 after B maximum light (plus upper
limits on day~+624). We have corrected all our photometric
measurements for the deviation of each instrumental photometric
setup from the Bessell (optical) and Persson (IR) standard
systems. This was done using the S-correction method, and was
applied up to the latest observed phases.

Besides the atypically red observed colours, evidence for high
extinction towards SN 2003cg include (a) its coincidence with a dust
lane of NGC 3169, (b) the very strong \NaI D interstellar doublet and
(c) the presence of a \textit{diffuse interstellar band}
(DIB). However, dereddening using a standard extinction law to match a
typical SN~Ia $B-V$ colour yields a peak absolute magnitude at maximum
which is abnormally bright. We therefore allowed \rv~to become a free
parameter within the CCM extinction law and adjusted \rv~and $E(B-V)$
to provide simultaneous matches to a range of colour curves of normal
SNe~Ia. From this, we obtained \ebv\, = 1.33 $\pm$ 0.11 and \rv = 1.80
$\pm$ 0.20. While the value obtained for \rv~is small, such values
have been invoked in the past. It implies that the grain size is small
compared with the average value for the local ISM. As an alternative
explanation, the light echo (LE) hypothesis \citep{wangl05}, has
severe difficulties. It seems unlikely that dust as close as
$R_0\leq$10$^{16}$cm could survive the supernova peak luminosity.  If
a significant proportion did survive it would produce a strong NIR
excess, but this is not seen.  Other LE effects expected but not seen
in SN~2003cg, include (a) a reduced $(B-V)$ color range, (b) temporal
variation in the reddening, (c) an anomalously small $\Delta m_{15}$,
(d) a significantly brighter late phase tail and (e) broader spectral
lines (more details in \citealt{patat05}). \\

The shape of the UBVRIJHK light curves of SN~2003cg are generally
typical of a ´normal´ Type Ia SN. The U and V light curves show a
broader peak and a pronounced shoulder in the post-maximum decay,
respectively. Again we believe that this is due to the high
reddening, causing the effective $\lambda$ of these bandpasses to
shift to redder wavelengths. We obtain a reddening-corrected
$\Delta m_{15}(B)_{obs}$ = 1.25 $\pm$ 0.05 for SN~2003cg, which is
typical for normal SNe~Ia. The intrinsic peak
bolometric luminosity is log$_{10}L$ = 43.01 $\pm$ 0.05 \ergs.  \\

The spectral evolution of SN 2003cg is also similar to that of other
normal Type Ia SN. This includes the spectral features, their ratios
and their evolution. The earliest spectra exhibit a weak, but clearly
visible, high velocity component of \CaII~IR. The late phase spectra
show emission features of [\FeII], [\FeIII] and [\CoII] plus an
unusual blend of lines between 7100 and 7350 $\AA$.  From the velocity
evolution of \SiII\, at 6355 $\AA$ and \SII\, at 5640 $\AA$, we
classify SN 2003cg as an LVG SN.  The same conclusion is indicated by
the $\cal R$(SiII) evolution.\\

SN~2003cg, the fourth target followed by the ESC, has provided an
important addition to the database of well-monitored nearby Type Ia.
In addition, further study of the heavy reddening towards SN~2003cg
will help us to establish the diversity in the characteristics of dust
responsible for astrophysical extinction. \\

\bigskip

\noindent {\bf ACKNOWLEDGMENTS}

We are grateful to the Visiting Astronomers of all telescopes who
kindly observed SN 2003cg as a \textit{Target of Opportunity}.
We thank also J. M\'endez's help in acquiring the TNG+OIG
observations, V. Stanishev for providing us his data of SN 2003du
prior to publication and J. A. Caballero for providing us a macro
for NIR data reduction.
This work is supported in part by the European Community's Human
Potential Programme under contract HPRN-CT-2002-00303, ``The
Physics of Type Ia Supernovae''.
This work is based on observations collected at the European
Southern Observatory, Chile under programmes ID 071.A-9006(A), ID
071.D-0281(D), ID 073.D-0853(A,B,C,D), 074.D-0340(E), at the Calar
Alto Observatory (Spain), the Italian Telescopio Nazionale Galileo
(La Palma), the Isaac Newton and William Herschel Telescopes of
the Isaac Newton Group (La Palma), the Asiago Observatory (Italy),
the Beijing Observatory (China), the Siding Spring Observatory
(Australia) and the United Kingdom Infrared Telescope (Hawaii).
The TNG is operated on the island of La Palma by the Centro
Galileo Galilei of INAF (Istituto Nazionale di Astrofisica) at the
Spanish Observatorio del Roque de los Muchachos of the Instituto
de Astrofisica de Canarias. The INT, JKT and WHT are operated on
the island of La Palma by the Isaac Newton Group (ING) in the
Spanish Observatorio del Roque de los Muchachos of the Instituto
de Astrofisica de Canarias. UKIRT is operated by the Joint
Astronomy Centre on behalf of the U.K. Particle Physics and
Astronomy Research Council. Some of the data reported here were
obtained as part of the ING and UKIRT Service Programmes.
This work has made use of the NASA/IPAC Extragalactic Database
(NED) which is operated by the Jet Propulsion Laboratory,
California Institute of Technology, under contract with the
National Aeronautics and Space Administration.

\end{document}